%% file: Vol2_ch6_master.tex
\documentclass[12pt]{article}

\usepackage[margin=1in]{geometry}
\usepackage{setspace}
\usepackage{amsmath,amssymb,amsthm}
\usepackage{bm}
\usepackage[hidelinks]{hyperref}

\usepackage{comment}

\usepackage{natbib}
\setcitestyle{authoryear,open={(},close={)},aysep={}}

\usepackage{graphicx}
\usepackage{tikz,pgfplots}
\pgfplotsset{compat=newest}

\usepackage{amsthm}

\numberwithin{equation}{section}

\theoremstyle{plain} 

\newtheorem{proposition}{Proposition}[section]
\newtheorem{lemma}{Lemma}[section]
\newtheorem{corollary}{Corollary}[section]

\theoremstyle{definition} 
\newtheorem{definition}{Definition}[section]
\newtheorem{assumption}{Assumption}[section]

\theoremstyle{remark} 

\newtheorem{example}{Example}[section]

\newtheorem*{theorem*}{Theorem}
\newtheorem*{proposition*}{Proposition}
\newtheorem*{lemma*}{Lemma}
\newtheorem*{corollary*}{Corollary}
\newtheorem*{definition*}{Definition}
\newtheorem*{assumption*}{Assumption}
\newtheorem*{remark*}{Remark}
\newtheorem*{example*}{Example}

\newcommand{\EE}{\mathbb{E}}

\newcommand{\KL}{\mathrm{KL}}
\newcommand{\Thetam}{\Theta^{\mathrm m}}

\title{Learning and Equilibrium under Model Misspecification 
\thanks{We thank Mira Frick, Drew Fudenberg, Ryota Iijima, Yuhta Ishii, Ran Spiegler, Philipp Strack, Yuichi Yamamoto, an anonymous referee, and especially Giacomo Lanzani for helpful comments. Esponda: iesponda@ucsb.edu. Pouzo: dpouzo@berkely.edu.}}

\author{Ignacio Esponda \and Demian Pouzo}
\date{\today}

\begin{document}
\maketitle
\begin{abstract}
This chapter develops a unified framework for studying misspecified learning situations in which agents optimize and update beliefs within an incorrect model of their environment. We review the statistical foundations of learning from misspecified models and extend these insights to environments with endogenous, action-dependent data, including both single agent and strategic settings. 
\end{abstract}

\begin{center}
    {\it{``All models are wrong, but some are useful" George Box.}}
\end{center}


\input{Vol2_ch6_intro}

\input{Vol2_ch6_exogenous}

\input{Vol2_ch6_single}
\input{Vol2_ch6_game}

\input{Vol2_ch6_convergence}

\bibliography{Vol2_ch6_references.bib}

\end{document}

%% file: Vol2_ch6_intro.tex
\section{Introduction}

Situations in which agents hold incorrect beliefs about their environment are ubiquitous in economics. Such belief distortions can arise through two broad channels: behavioral mechanisms that systematically bias belief formation (e.g., overconfidence, correlation neglect, omitted-variable bias), and pragmatic simplifications whereby agents deliberately adopt reduced models because the true environment is too complex to represent, or only coarse, limited, signals are observable, or there is some degree of bounded-rationality. This chapter develops a unified conceptual framework for misspecified learning that encompasses both sources of misspecification within a single analytical lens. In this framework, agents are modeled as optimizing within a constrained class of subjective models. Rather than specifying beliefs exogenously, the framework delineates the model class, the information structure, and the updating rule through which agents interpret data; beliefs then emerge endogenously from this learning process, and actions are determined optimally given those beliefs.

Although we focus on a particular class of repeated environments, the underlying method is more general. The framework is designed to make explicit what the environment is, what agents do and do not observe, and which (possibly misspecified) model they use to interpret the data. Given these primitives and the feedback generated along the path of play, beliefs and behavior are endogenously determined. We view this discipline as the main value of the approach and as a template for extensions beyond the settings covered in this chapter.

We present a self-contained exposition of misspecified learning in both single-agent and strategic settings. We lay out the primitives---the agent's subjective model class, the data structure, and the updating rule---and show how these generate beliefs and choices over time. We discuss an equilibrium notion tailored to misspecification (Berk--Nash equilibrium) and survey methods for proving convergence to equilibrium (or demonstrating nonconvergence) and for bounding long-run play. Our style is to state formal definitions and propositions, provide proof sketches that emphasize the key ideas and intuition, and point to the original sources for full arguments; along the way, we highlight extensions and variants developed in the literature.

This chapter is not a comprehensive survey of misspecification in economics. We set aside the macroeconomic literature on learning under misspecification (e.g., \cite{evans2001learning}, \cite{branch2006restricted}, \cite{hansen2011robustness})
and we also omit much of the microeconomic literature on misspecification, which is surveyed in \cite{bohren2025misspecified}. Our aim is complementary: to develop a narrower set of ideas in depth, unify notation, and highlight the main proof strategies to clarify when and why the results apply. The references we provide point to the original papers and to further literatures with more extensive bibliographies for the interested reader.


Section~\ref{sec:exo} begins with learning from \emph{exogenous} data---the classical setting studied by statisticians. Sections~\ref{sec:single}–\ref{sec:conv} then turn to learning from \emph{endogenous} data, where one or more agents (or players) choose actions and thereby shape the data they observe---a line of work developed primarily in economics. While many action environments could be analyzed, we focus on a simple and standard repeated setting in which the same decision problem recurs each period, using it to illustrate the main ideas. Misspecified learning has also been explored in other classes of problems---including a sizable literature on social learning in which agents draw (possibly incorrect) inferences from predecessors’ actions (e.g., \cite{eyster2010naive,bohren2016informational,bohren2021learning,frick2020misinterpreting}), Markov decision processes (\cite{esponda2021equilibrium,anderson2024existence}), optimal stopping (\cite{he2022mislearning}), and recursive macroeconomic models (e.g., \cite{molavi2019macroeconomics,FusterEtAlJEP2010}), among others. These settings borrow some of the tools developed for our main environment, but we do not cover them in this chapter.

We conclude the introduction by situating this chapter's topic in a broader context---briefly tracing the intellectual path that led to the current framework and pointing to recent development and directions for future work---so as to clarify how we arrived here and describe where the literature seems to be going next.

\subsection{Five strands in the development of misspecified learning in economics}

In this subsection, we situate the chapter within a broader historical arc by sketching five strands that shaped the modern framework. 
The goal is not completeness (we omit several important contributions) but orientation: a roadmap that clarifies how the field arrived at the current toolkit and where recent work is heading.

A first strand dates to the 1970s and links learning to equilibrium. \cite{hahn1973notion}’s view was that equilibrium requires “learning to be exhausted”: incoming data no longer lead agents to revise their  theory. In unpublished work, \cite{ArrowGreen1973-IMSSS} laid out an early framework that can be read as misspecification: they asked when an equilibrium could exist, what properties it would have, and whether it could be reached by a learning process starting from nonequilibrium beliefs, with the requirement that the realized \emph{distribution} of outcomes match the \emph{anticipated} distribution---not merely its mean. \cite{kirman1975learning} made the misspecification angle explicit by modeling firms that hold erroneous theories of their environment. What these early contributions lacked, however, was a systematic connection to the then-contemporary statistical literature on learning from misspecified models---most notably \cite{Berk1966}.

A second strand, emerging in the late 1980s and early 1990s, develops the notion of \emph{self-confirming equilibrium} (SCE). In Nash equilibrium, players’ beliefs are correct both on and off the equilibrium path; in SCE, by contrast, beliefs need only be correct \emph{on the path of play}. More precisely, each player’s strategy is optimal given their beliefs, and those beliefs are consistent with the feedback generated by the realized play, but beliefs about unreached contingencies (off-path information sets) need not be correct. This captures one reading of Hahn’s “no further learning” idea: incoming data confirm the agent’s theory in the sense that nothing observed contradicts it, without implying the theory is globally correct. \cite{fudenberg1993self} develop SCE in games (see also \cite{Battigalli1987} and \cite{fudenberg1988learning}) and, together with contemporaneous work on adaptive learning (e.g., \cite{fudenberg1993steady, fudenberg1995learning}), analyze when Nash or SCE can arise as long-run outcomes of learning dynamics---see \cite{fudenberg1998theory} for a textbook treatment of learning in games. Importantly, SCE is \emph{not} inherently about misspecification: agents can possess a correct model of the game, yet hold incorrect beliefs about off-path play simply because those contingencies never generate data. In misspecified learning, by contrast, the agent’s \emph{model class} itself excludes the true data-generating process, so even on-path feedback may be best rationalized by an incorrect theory.

A third strand, emerging in the early 2000s, introduces equilibrium notions that embed specific incorrect beliefs. \emph{Cursed equilibrium} and \emph{behavioral equilibrium} capture neglect of the informational content of others' actions \citep{eyster2005cursed,esponda2008behavioral}. \emph{Analogy-based expectation equilibrium} formalizes coarse reasoning: agents cluster decision situations into a few \emph{analogy classes}, form class-level expectations, and best-respond to them; these expectations must be correct on average within each class \citep{jehiel2008revisiting}. These are special cases of misspecifications, and they are also \emph{static}: the biases are imposed in the equilibrium definition rather than explicitly derived from a data-generating process and learning rule, and they do not yet engage with the statistical theory of misspecified learning.

A fourth strand recognizes the value of modeling diverse forms of misspecification within a common framework and explicitly imports tools from the statistics of misspecified learning to develop both general solution concepts and learning foundations. This is the body of work we review in this chapter. The timing was natural: concerns about incorrect models---though seldom labeled “misspecification” in earlier behavioral and bounded-rationality literatures---had reemerged, while the learning-in-games literature offered a route to formal foundations but lacked a tight link to statistical results on asymptotic belief characterization under misspecified models. The key step was to adapt those statistical insights---originally derived by statistician for exogenous data---to environments where data are \emph{endogenous}, shaped by agents' actions. The picture that emerges is clear: whereas learning-based equilibrium (Nash and self-confirming) emphasizes that beliefs are restricted by what agents \emph{observe}, it is equally important that beliefs are shaped by the \emph{model} through which observations are interpreted. 

A fifth strand is now emerging. The work reviewed in this chapter takes agents' (possibly misspecified) models as \emph{exogenous}: by assumption, observed evidence never directly falsifies the model, and agents do not entertain the possibility that their model is wrong. This can be natural---for example, when formalizing a behavioral bias that is unrecognized by the agent---but in many settings agents are aware that their models are approximations. This raises at least two related questions. \emph{First}, how are models selected and how do they evolve over time—i.e., which misspecifications are likely to persist, and why? \emph{Second}, what is the appropriate positive and normative decision-theoretic framework for agents who acknowledge possible misspecification and wish to guard against it (e.g., via robust decision rules)? An active literature is beginning to address these issues; due to space constraints we do not cover these developments here (for references, see \cite{bohren2025misspecified} for the first question and \cite{lanzani2025dynamic} for the second one).

%% file: Vol2_ch6_exogenous.tex
\section{Learning from exogenous data} \label{sec:exo}

Consider learning from a sequence of exogenous observations
\(Y_1,Y_2,\ldots\) that are IID under a true law \(Q \) and each observation lies in a Borel subset \((\mathbb{Y},\mathcal{Y})\) of a Euclidean space.
The agent entertains an IID \emph{subjective model} over the observation \(Y_t \) given by \(\mathcal{Q}(\Theta) : = \{Q_\theta:\theta\in\Theta\}\). This model can be correctly specified, $Q \in \mathcal{Q}(\Theta)$, or misspecified, $Q \notin \mathcal{Q}(\Theta)$.


We assume there exists a $\sigma$-finite measure $\nu$ dominating both $Q$ and all $Q_\theta$, with densities $q$ and $q_\theta$. Leaving $\nu$ unspecified accommodates both continuous (Lebesgue) and discrete (counting) settings.

Given data \(Y_1,\ldots,Y_t\), the \emph{Bayesian posterior} is denoted by
\(\mu_t(\cdot) : = \mu(\cdot\mid Y_1,\ldots,Y_t)\), and by the Bayesian updating rule and some simple algebra, is given (when it exists) by \footnote{Below we provide asssumptions that ensure the existence of the Bayesian posterior.}
\begin{align}\label{eqn:bayes.update}
    \mu_{t}(B) = \frac{\int_{B} e^{- \sum_{i=1}^{t} \log \frac{q(Y_{i})}{q_{\theta}(Y_{i})} } \mu_{0}(d \theta) }{\int_{\Theta} e^{ - \sum_{i=1}^{t} \log \frac{q(Y_{i})}{q_{\theta}(Y_{i})}  } \mu_{0}(d\theta)   },\text{ for any Borel set $B \subseteq \Theta$,} 
\end{align}
where $\theta \mapsto - \sum_{i=1}^{t} \log \frac{q(Y_{i})}{q_{\theta}(Y_{i})}$ is the log-likelihood ratio function given sample $(Y_{1},...,Y_{t})$.

For reference, the true probability distribution generates a probability measure over infinite histories $(Y_{1},Y_{2},\ldots)$ denoted by $\mathbf P$. Analogously, each element of the subjective model $\mathcal{Q}(\Theta)$ generates a subjective probability measure over infinite histories denoted by $\mathbf P_{\theta}$, and the prior also induces a subjective probability measure given by \( \int_{\Theta} \mathbf P_\theta \mu_0(d\theta)\).

\subsection{Posterior Concentration}

A desirable property of Bayesian updating is that, with enough data, agents eventually learn the truth --- or, under miss-specification, at least an approximation to it. To formalize this idea, the literature studies the asymptotic concentration of the posterior distribution. The main result in this section establishes such convergence.

To achieve such results we need the following regularity conditions 

\begin{assumption}[Parameter space]\label{ass:Theta.compact}
    The parameter set,  $\Theta$, is a compact subset of some Euclidean space.
\end{assumption}

\begin{assumption} [Domination and a.e.\ continuity] \label{ass:PDF.dominance} For \(\nu\)-a.e.\ \(y\), the map \( \theta \mapsto q_\theta(y)\) is continuous. There exist measurable \(M:\mathbb{Y}\to[1,\infty)\) such that, for all \(\theta\) and \(\nu\)-a.e.\ \(y\),
\[
q_\theta(y)\,M(y)^{-1}\ \le\ q(y)\ \le\ q_\theta(y)\,M(y)~and~\int M(y) q(y) \nu(dy)<\infty.
\]
\end{assumption}

\begin{assumption}[Full support]\label{ass:prior.supp}
  The prior density is bounded away from zero on \(\Theta\). 
\end{assumption}

\paragraph{Doob's Martingale Convergence Theorem.} For each Borel set \(B \subseteq \Theta\), the sequence \((\mu_t(B))_{t\ge 0}\) is a bounded
martingale with respect to the filtration generated by the data under the measure \(\int_{\Theta} \mathbf P_\theta \mu_0(d\theta) \). Thus, by Doob’s martingale convergence theorem (\cite{doob1953stochastic}), \(\mu_t(\theta)\) converges \(\int_{\Theta} \mathbf P_\theta \mu_0(d\theta)\)-almost surely as \(t\to\infty\). Since any event \(H\) that satisfies \(1=\int_\Theta \mathbf P_\theta(H)\,\mu_0(d\theta)\) implies \(\mathbf P_\theta(H)=1\) for \(\mu_0\)-a.e.\ \(\theta\), this convergence result under \(\int_{\Theta} \mathbf P_\theta \mu_0(d\theta)\) transfers to \(\mathbf P_\theta\) for $\mu_{0}$-a.e. \(\theta\).

This result however, has a few important limitations. First, the convergence provided by Doob's martingale theorem holds only 
for $\mu_0$-almost every parameter values. Hence, even in the correctly specified case, where there exists a $\theta^{\star} \in \Theta$ such that $\mathbf P = \mathbf P_{\theta^\star}$, it may
happen that the true parameter $\theta^\star$ lies in the exceptional $\mu_0$-null set where the theorem is silent. Second, and more importantly for our purposes, the argument does not apply to the misspecified case, since the relevant probability law is $\mathbf P$ rather than the law $\int_{\Theta} \mathbf P_\theta \mu_0(d\theta)$ under which the martingale convergence is derived. 

For these reasons, standard Bayesian consistency theory relies instead on a different organizing principle based on frequentist-probability arguments (e.g. \cite{Schwartz1965}), which establish convergence under $\mathbf P $ and thus guarantee posterior concentration for the true data-generating distribution rather than merely for $\mu_0$-almost every parameter value.

A crucial concept within this approach is to construct a ``natural notion of distance" between the agent's model and the true probaility distribution. As it turns out this notion of distance is captured by the Kullback--Leibler (KL) divergence. 

\paragraph{Kullback--Leibler (KL) divergence.} For any $\theta \in \Theta$, the KL divergence between $Q_\theta$ and $Q$ is given by
\[
\KL(Q^\star\Vert Q_\theta)\ :=\ \EE_{Q} \!\left[\log\frac{q (Y)}{q_\theta(Y)}\right]\ \in[0,\infty].
\]
We define the \emph{KL-minimizing set}, which is correctly defined under our conditions, as 
\[
\Thetam\ :=\ \arg\min_{\theta\in\Theta}\ \KL(Q \Vert Q_\theta).
\]

\paragraph{Asymptotic Concentration of Beliefs.} 
The next result identifies the Bayesian posterior’s asymptotic support under the true law $\mathbf P$, providing a quantitative version of \cite{Berk1966}'s classical theorem.

\begin{proposition}\label{pro:Berk}
For all closed set $C \subset \Theta \setminus \Theta^{m}$, there exists a constant $L<\infty$ such that \footnote{The symbol $o_{as}(1)$ denotes a positive random variable that $\mathbf P$-a.s. converges to zero.}
\begin{align}
    \mu_{t}(C) \leq L e^{-t (\rho_{C} - o_{as}(1)) },~\forall t \geq 0,
\end{align}
where $\rho_{C} : = \min_{\theta \in C} \KL(Q \Vert Q_\theta) - \min_{\theta \in \Theta} \KL(Q \Vert Q_\theta) > 0$. 
\end{proposition}

\paragraph{Discussion.} This exponential decay of posterior mass on a ``bad'' set $C$ follows from 
\eqref{eqn:bayes.update} and three key insights. First, the log-likelihood ratio is 
asymptotically driven by the negative Kullback--Leibler divergence, so parameters with larger 
KL divergence are down–weighted. Second, $C$ is ``well separated'' from 
$\Theta^{m}$, in that its KL gap $\rho_{C}$ is strictly positive, giving an exponential bound 
on the numerator of \eqref{eqn:bayes.update}. Third, the prior assigns positive mass to every neighborhood of $\Theta^{m}$, ensuring that the denominator in \ref{eqn:bayes.update} remains of the correct exponential order and does not collapse. 

\begin{proof}[Sketch of the Proof of Proposition \ref{pro:Berk}]
    By expression \ref{eqn:bayes.update} the concentration of the posterior belief is dictated by the log-likelihood ratio function $\theta \mapsto \sum_{i=1}^{t} \log \frac{q_{\theta}(Y_{i})}{q(Y_{i})}$.

By the Law of Large Numbers (LLN), for each $\theta \in \Theta$, the (scaled) log-likelihood ratio function converges to $-\EE_{Q} \left[ \log \frac{q(Y)}{q_{\theta}(Y)}  \right]$. Under our conditions, the family $\left\{ \theta \mapsto t^{-1} \sum_{i=1}^{t} \log \frac{q_{\theta}(Y_{i})}{q(Y_{i})} \colon t \in \mathbb{N}   \right\} $ is stochastic equi-continuous (cf. \cite{newey1994large}) and thus by a version of the Arzela-Ascoli Theorem the convergence is extended to hold uniformly over $\Theta$, i.e.,
\begin{align*}
    \sup_{\theta \in \Theta} \left|t^{-1} \sum_{i=1}^{t} \log \frac{q_{\theta}(Y_{i})}{q(Y_{i})} + KL(Q \mid \mid Q_{\theta} ) \right| = o_{as}(1).
\end{align*}

Therefore, for any $C \subseteq \Theta$ Borel, 
\begin{align*}
    \mu_{t}(C) \leq  \frac{\int_{C} e^{t( - KL(Q \mid \mid Q_{\theta} ) + o_{as}(1) )} \mu_{0}(d \theta) }{\int_{\Theta} e^{t ( - KL(Q \mid \mid Q_{\theta} ) - o_{as}(1)  ) } \mu_{0}(d\theta)   }   =  \frac{\int_{C} e^{-t( KL(Q \mid \mid Q_{\theta} ) - KL^{m} + o_{as}(1)  )} \mu_{0}(d \theta) }{\int_{\Theta} e^{-t ( KL(Q \mid \mid Q_{\theta} ) - KL^{m} - o_{as}(1)  ) } \mu_{0}(d\theta)   }
\end{align*}
where $KL^{m} : = \min_{\theta \in \Theta} KL (Q \mid \mid Q_{\theta} )$.

Consider now a set $C \subseteq \Theta$ that is closed and $C \cap \Theta^{m} = \{\emptyset\}$. Since both $C$ and $\Theta^{m}$ are compact and disjoint, the KL divergence over $C$ is ``well separated" from the set of minimizers $\Theta^{m}$, i.e., $\rho_{C} : =  \inf_{\theta \in C} KL(Q \mid \mid Q_{\theta} ) - KL^{m}  > 0.$
Therefore, $\int_{C} e^{-t( KL(Q \mid \mid Q_{\theta} ) - KL^{m} + o_{as}(1)  )} \mu_{0}(d \theta) \leq \int_{C} e^{-t ( \rho_{C} + o_{as}(1)  )} \mu_{0}(d \theta) \leq e^{-t ( \rho_{C} + o_{as}(1)  )} $. 

Moreover, for any such $\rho_{C} > 0$, the set $\Theta_{\rho_{C}} : = \{ \theta \in \Theta \colon KL(Q \mid \mid Q_{\theta} ) - KL^{m} < 0.5 \rho_{C} \} $ has non-empty interior --- is the pre-image of an open set under a continuous function ---, and under full-support condition has positive probability under the prior. Therefore, for any set $C$ that is closed and $C \cap \Theta^{m} = \{\emptyset\}$, 
\begin{align*}
    \mu_{t}(C) \leq  \frac{e^{-t ( \rho_{C} + o_{as}(1)  )}   }{\int_{\Theta_{\rho_{C}}} e^{ -t ( KL(Q \mid \mid Q_{\theta} ) - KL^{m} - o_{as}(1)  ) } \mu_{0}(d\theta)   }   \leq  \frac{e^{-t ( \rho_{C} + o_{as}(1)  )}   }{ e^{- t ( 0.5 \rho_{C} - o_{as}(1)  ) }  \mu_{0} (\Theta_{\rho_{C}})} \leq \frac{e^{-t ( 0.5 \rho_{C} - o_{as}(1)  )}   }{ \mu_{0}(\Theta_{\rho_{C}})}.
\end{align*}

Thus, the desired result is achieved with the constant given by $1/\mu_{0}(\Theta_{\rho_{C}})$.
\end{proof}

\paragraph{Implications of Proposition \ref{pro:Berk}.} 
If a unique parameter minimizes the KL divergence, the posterior converges and 
concentrates at this point.\footnote{$\Rightarrow$ denotes weak convergence of probability 
measures; $\delta_{\theta}$ is the point mass at $\theta$.} Formally,
\begin{corollary}
    If \(\Thetam=\{\theta^{\mathrm m}\}\), then \(\mu_t\Rightarrow\delta_{\theta^{\mathrm m}}\) \(\mathbf P\)-a.s.
\end{corollary}

In particular, Corollary \ref{pro:Berk} readily implies that the agent will eventually learn the true distribution  generating the data if the model is correctly specified.

Finally, if $\Theta^{m}$ is not a singleton, then Bayesian posterior need not converge to a point. At this level of generality, one can only conclude that it will eventually concentrate on the set of best-fit parameters, $\Theta^{m}$.

\begin{example}[Coin Flips (Bernoulli model)]
Let \(Y\in\{0,1\}\) and suppose the model posits \(Q_\theta(Y=1)=\theta\) with \(\Theta\subseteq [0,1] \).
The Kullback--Leibler divergence between a true success probability \(\theta^\star\) and a model \(\theta\) is
\[
\KL( \theta^\star\Vert \theta) : = KL(Q_{\theta^{\star}} \Vert Q_{\theta})
\;=\;
\theta^\star \log\!\Big(\frac{\theta^\star}{\theta}\Big)
\;+\;
(1-\theta^\star)\log\!\Big(\frac{1-\theta^\star}{1-\theta}\Big).
\]
This example is useful to visualize two facts:
(i) \(\KL(\theta^\star\Vert \theta)\) is generally \emph{asymmetric} in its arguments
(\(\KL(\theta^\star\Vert \theta)\neq \KL(\theta\Vert \theta^\star)\)); and
(ii) it is not a metric and does not behave like Euclidean distance on \((0,1)\)
(the level sets are skewed toward the boundaries, reflecting likelihood curvature).

\begin{figure}[h!]
  \centering
  \begin{tikzpicture}
    \begin{axis}[
      xlabel={$\theta$}, ylabel={$\KL$},
      xmin=0, xmax=1, ymin=0, ymax=1.5,
      xtick={0,0.25,0.5,0.6,0.75,0.9,1},
      ytick={0,0.5,1,1.5},
      legend pos=north east,
      grid=major,
      width=0.9\textwidth, height=7.5cm
    ]
      \addplot[domain=0.001:0.999, samples=300]
        {0.5*ln(0.5/x) + 0.5*ln(0.5/(1-x))};
      \addlegendentry{$\KL(0.5\Vert \theta)$}

      \addplot[domain=0.001:0.999, samples=300]
        {0.75*ln(0.75/x) + 0.25*ln(0.25/(1-x))};
      \addlegendentry{$\KL(0.75\Vert \theta)$}
    \end{axis}
  \end{tikzpicture}
  \caption{Bernoulli KL curves for two truths, \(\theta^\star=0.5\) and \(\theta^\star=0.75\).}
  \label{fig:bern-kl}
\end{figure}
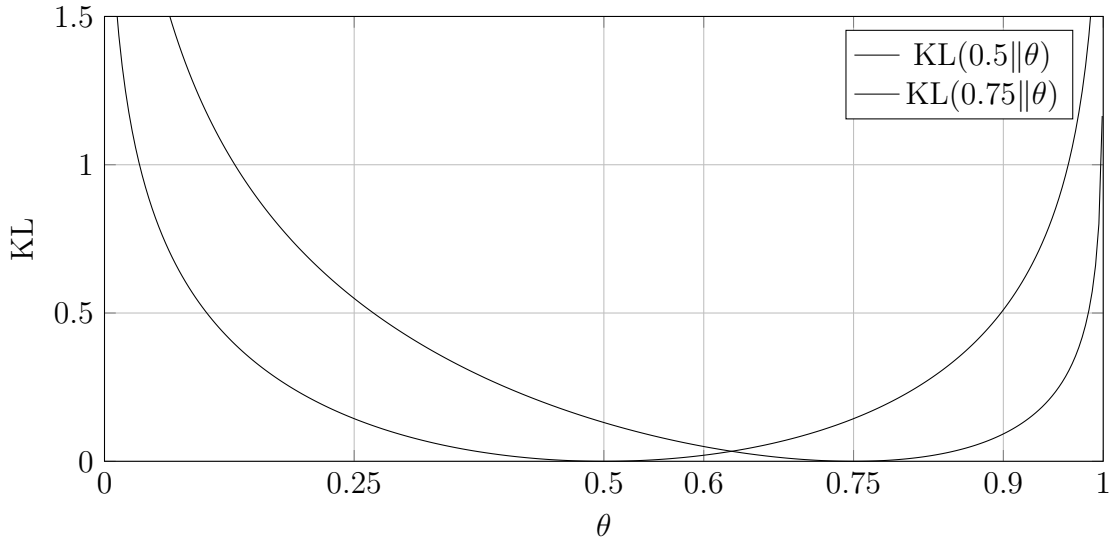

In order to characterize $\Theta^{m}$, it is easy to see that $\theta \mapsto KL(\theta \Vert \theta^{\star})$ is decreasing for $\theta < \theta^\star$ and increasing for $\theta > \theta^\star$. Therefore, for the correctly specified case where $\Theta \ni \theta^{\star}$, the  $\theta^{\star}$ is the global minimum and thus $\Theta^{m} = \{\theta^{\star}\}$. On the other hand, if the agent's model is misspecified, i.e., $\theta^{\star} \notin \Theta$, the situation is qualitatively different. Indeed $\Theta^{m}$ is comprised of, at most, two elements: The ``projections" of $\theta^{\star}$ onto $\Theta$ from the right and the left. For instance, if $\theta^{\star} = 0.50$ then $\Theta^{m}$ may consist of just one element --- the projection that is closest to the true parameter $\theta^{\star}$. So in this case, while the model is misspecified, we still have a unique minimizer that represents the best fit within the model. On the other hand, if the projections are equi-distant to the true parameter, $\Theta^{m} $ is comprised of these two elements. That is, there are two parameters that fit the data equally well under the KL divergence. 

This difference has important consequences for the concentration of the Bayesian posterior. In the correctly specified case, since \(\Thetam= \theta^{\star} \),  the Bayesian posterior concentrates at the truth: \(\mu_t\Rightarrow\delta_{\theta^{\star}}\) \(\mathbf P \)-a.s. Concentration is retained in the misspecified case with a unique best fit parameter. For instance, if \(\theta^\star=0.5\) but
\(\Theta=\{0.25,0.3,0.6,0.9\}\), then \(\Thetam=\{0.6\}\) and
\(\mu_t\Rightarrow\delta_{0.6}\) \(\mathbf P \)-a.s. Thus, Bayesian posteriors still converge, but the limit parameter is the KL projection (pseudo-true value), not the truth. 

On the other hand, if there is non-uniqueness of the best fit parameters Bayesian posteriors may not converge. To see this, let \(\theta^\star=0.5\) and \(\Theta=\{0.25,\,0.75\}\). By our arguments above, $\Theta^{m} = \{0.25,0.75\}$. In this case the Bayesian posterior \(\mu_t(\cdot)\) \emph{does not converge}:
\[
\liminf_{t\to\infty}\mu_t(0.25)=0
\quad\text{and}\quad
\limsup_{t\to\infty}\mu_t(0.25)=1
\quad (\mathbf P \text{-a.s.}).
\]
Indeed, by Bayes’ rule,
\begin{align*}
    \log \frac{\mu_{t}(0.25)}{\mu_{t}(0.75)} = \log \frac{\mu_{0}(0.25)}{\mu_{0}(0.75)} + \sum_{i=1}^{t} \xi_i
\end{align*}
where \(\xi_i:=\log\!\big(Q_{0.25}(Y_i)/Q_{0.75}(Y_i)\big)\). Observe that \(\xi_i\) are IID with $\mathbb{E}[\xi_{i}] = 0$ (the two models are tied in KL). So the log likelihood ratio of posteriors is a zero-drift random walk.

Therefore, by the properties of a random walk, the $( \log \frac{\mu_{t}(0.25)}{\mu_{t}(0.75)} )_{t}$ will $\mathbf P$-a.s. attains arbitrarily large positive values and arbitrarily large negative values and will return to zero infinitely often. This translates that Bayesian posteriors will constantly oscillate between periods of assigning almost all the mass to $0.25$ to periods assigning almost all the mass to $0.75$, never settling down. Moreover, this result is independent of the particular prior. \(\blacklozenge\)

\end{example}

\subsection{Biographical Notes}

The study of Bayesian posterior concentration originates in the seminal work of \citet{Schwartz1965}, who established the first general consistency theorem for Bayesian models under correct specification. Schwartz showed that if the prior assigns positive mass to every Kullback--Leibler (KL) neighborhood of the true distribution and uniformly consistent tests exist separating the truth from its complement, then the posterior concentrates (in the weak topology) around the data-generating distribution. Schwartz's result made explicit that posterior consistency is driven not only by prior ``support'' but by a frequentist testing property, a perspective that subsequently became the foundation of modern Bayesian asymptotics.

\citet{Berk1966} extended this analysis to the \emph{misspecified} parametric case. When the data-generating distribution lies outside the assumed model, he proved that the posterior still concentrates, but now around the \emph{KL projection} of the truth---the element of the model minimizing $KL(Q \Vert Q_{\theta} )$. Thus, miss-specification affects the \emph{limit point} of posterior learning, but not the fact that the posterior concentrates. The non-convergence of beliefs in the coin tossing example is due to this paper. A more refined treatment of Berk's misspecified parametric case was later given by \citet{BunkeMilhaud1998}, who strengthened Berk’s conclusion by studying not only the posterior itself but also the asymptotic behavior of Bayes estimators under general loss functions. \cite{YAMADA1976294} also extends Berk's in many dimensions and provides a more modern, transparent, proof of the result, akin to the sketch of the proof provided here. \citet{ShaliziEJS2009} extends these results to dependent data by relying on the asymptotic equipartition (Shannon-McMillan-Breiman) property of information theory. Parallel to the Berk's results for misspecified models, \cite{white1982maximum} establishes frequentist results whereby he showed the maximum  (quasi) likelihood estimator that maximizes the average log-(quasi) likelihood \(\frac{1}{t}\sum_{i=1}^t \log q_\theta(Y_i)\) converges to \(\min_\theta \KL(Q \Vert Q_\theta)\). Thus, in IID exogenous settings, Bayesian and frequentist procedures target the same parameters: those that provide a best fit under the KL divergence.

The regularity conditions used in Proposition \ref{pro:Berk} are inspired by those in \citet{Schwartz1965}. The continuity and dominance conditions and compactness condition are required to establish uniform convergence of the log-likelihood ratio function to the KL divergence. 

By inspection of the proof of the proposition, it is apparent the finite dimensional requirement of the parameter space was not used. Indeed, \citet{Schwartz1965} results allow for infinite dimensional parameters.\footnote{Other more recent references for infinite dimensional spaces include \citet{GhosalGhoshvanderVaart2000}, \cite{ShenWasserman2001}; see \cite{GhosalVanDerVaart2017} for a more thorough discussion and additional references.} However, in infinite dimensional cases, the verification of the regularity assumptions is not trivial.

As pointed out in the seminal paper by \cite{DiaconisFreedman1986}, the ``prior support" condition of \citet{Schwartz1965} is far from automatic in infinite-dimensional models. The reason is that this condition requires the prior to put positive probability on open neighborhoods \emph{under the KL divergence} of the truth. However, their examples show that a prior may be weakly dense and yet fail to put \emph{any} mass on sufficiently small KL neighbourhoods of the true distribution.  In such cases the prior is said to be \emph{thin} at $\mathbf P$, and Bayesian updating cannot recover the truth, because the posterior is not allowed to accumulate likelihood weight in a KL ball around the true probability distribution. The contrasting notion of a \emph{thick} prior refers to one that assigns strictly positive mass to every KL neighborhood of the data-generating law.  The necessity of this condition was made precise by \citet{BarronSchervishWasserman1999}, who showed that---under mild regularity conditions---KL thickness is not merely a sufficient condition (as in Schwartz's theorem), but essentially a \emph{necessary} one for posterior consistency in nonparametric settings.  In this way, the \cite{DiaconisFreedman1986} examples clarified that Schwartz's support condition is mathematically sharp: nonparametric Bayesian consistency relies fundamentally on KL \emph{thickness}, and not merely on weak or topological support.

\cite{KleijnvanderVaart2006} extends Schwartz's results to misspecified models thereby generalizing Berk’s result to nonparametric models under misspecification. Unlike in the correctly specified case of \cite{Schwartz1965}, the bottleneck is not just the ``KL thickness" condition but the ``well separation" of the set of minimizers --- this well separation relies on the construction of uniformly consistent tests around the pseudo-true distribution, whose local geometry may be much harder to control.

%% file: Vol2_ch6_single.tex
\section{Endogenous learning: single agent}\label{sec:single}

\subsection{Setup and equilibrium}

A single agent chooses an action \(a\in\mathbb{A}\). Consequences take values in \(\mathbb{Y}\).
The true consequence kernel is \(Q:\mathbb{A}\to\Delta\mathbb{Y}\).
The agent entertains a (possibly misspecified) parametric model \(\{Q_\theta:\theta\in\Theta\}\),
with each \(Q_\theta:\mathbb{A}\to\Delta\mathbb{Y}\).
The per-period payoff is \(\pi:\mathbb{A}\times\mathbb{Y}\to\mathbb{R}\).

\paragraph{Assumptions (single agent).}
\begin{itemize}
\item \emph{Compactness.} \(\mathbb{A}\) and \(\Theta\) are nonempty and compact subsets of Euclidean spaces; \(\mathbb{Y}\) is a Borel subset of a Euclidean space.
\item \emph{Domination and a.e.\ continuity of kernels.} There exists a \(\sigma\)-finite Borel measure \(\nu\) on \(\mathbb{Y}\) such that \(Q(\cdot\mid a)\ll\nu\) and \(Q_\theta(\cdot\mid a)\ll\nu\) for all \((\theta,a)\), with densities \(q(\cdot\mid a)\) and \(q_\theta(\cdot\mid a)\). For \(\nu\)-a.e.\ \(y\), the maps \(a\mapsto q(y\mid a)\) and \((\theta,a)\mapsto q_\theta(y\mid a)\) are continuous.
\item \emph{Likelihood ratio envelope.} There exist measurable \(M:\mathbb{Y}\to[1,\infty)\) and \(r:\mathbb{Y}\to[0,\infty)\) such that, for all \((\theta,a)\) and \(\nu\)-a.e.\ \(y\),
\[
q_\theta(y\mid a)\,M(y)^{-1}\ \le\ q(y\mid a)\ \le\ q_\theta(y\mid a)\,M(y),\qquad q(y\mid a)\le r(y),
\]
and \(\int M(y)\,r(y)\,\nu(dy)<\infty\).
\item \emph{Domination and a.e.\ continuity of the payoff.} For \(\nu\)-a.e.\ \(y\), \((a,y)\mapsto \pi(a,y)\) is continuous; there exists \(h_\pi:\mathbb{Y}\to[0,\infty)\) with \(|\pi(a,y)|\le h_\pi(y)\) for all \(a\) and \(\int h_\pi(y)\,M(y)\,r(y)\,\nu(dy)<\infty\).
\end{itemize}

These conditions accommodate discrete or continuous \(\mathbb{A}\) and \(\mathbb{Y}\). They ensure the existence of the set of minimizers of the Kullback–Leibler divergence and of the optimal action correspondence, and they also provide the uniform laws of large numbers required for the dynamic environment presented below. We maintain these standing assumptions throughout the chapter unless otherwise noted.

\paragraph{Optimality correspondence.}
For \(\theta\in\Theta\), define the expected payoff of action $a$
\[
U(a,\theta)\;:=\;\int_{\mathbb{Y}}\pi(a,y)\,Q_\theta(dy\mid a).
\]
Given a belief \(\mu\in\Delta\Theta\), the optimal action correspondence is
\[
F(\mu)\;:=\;\arg\max_{a\in\mathbb{A}} \int_{\Theta} U(a,\theta)\,\mu(d\theta).
\]
Under the standing assumptions, $F$ is nonempty, compact valued, and it is upper hemicontinuous (UHC) in $\mu$. Moreover, the results below apply to any action correspondence with these properties. Thus, while the misspecified-learning literature typically derives $F$ from maximization, the analysis also accommodates non-optimal choice rules.

\paragraph{Action-dependent KL divergence.}
The Kullback--Leibler divergence between the true kernel and a model \(Q_\theta\) at action \(a\) is
\[
K(\theta,a)\;:= KL(Q(\cdot \mid a) \Vert Q_{\theta}(\cdot \mid a) )  =  \int_{\mathbb{Y}}
\log\!\Big(\tfrac{q(y\mid a)}{q_\theta(y\mid a)}\Big)\,q(y\mid a)\,\nu(dy).
\]
For any action distribution \(\sigma\in\Delta\mathbb{A}\), define the KL-minimizing set
\[
\Theta^{\mathrm m}(\sigma)\;:=\;\arg\min_{\theta\in\Theta}\ \int_{\mathbb{A}} K(\theta,a)\,\sigma(da).
\]
Intuitively, \(\Theta^{\mathrm m}(\sigma)\) collects the parameter  values whose kernels best fit the true data induced by long-run action frequencies \(\sigma\). Under our assumptions, $\Theta^{\mathrm m}$ is nonempty and compact valued and it is UHC in $\sigma$.

The goal is to understand what happens when the agent repeatedly faces this decision problem and updates her belief about the parameter space. 
We now introduce an equilibrium concept and, in Section~\ref{sub:dynamics}, formally develop the dynamic framework showing that equilibrium represents the steady state of this learning process. 
Actions determine the feedback the agent observes; that feedback determines her beliefs; and---under misspecification---those beliefs typically depend on how actions are sampled, generating a feedback loop between actions and beliefs. The steady state can thus be characterized as a fixed point of this process, even in a single-agent setting.

\begin{definition} \label{def:BeNE}
A probability measure \(\sigma\in\Delta \mathbb{A}\) is a \textbf{Berk--Nash equilibrium} of the single-agent problem if there exists a belief \(\mu\in\Delta \Theta\) such that:
\begin{enumerate}
\item \emph{Optimality:} \(\operatorname{supp}(\sigma)\subseteq F(\mu)\).
\item \emph{Belief restriction:} \(\operatorname{supp}(\mu)\subseteq \Theta^{\mathrm m}(\sigma)\).
\end{enumerate}
\end{definition}

\bigskip

\begin{example}{Returns to effort (overconfidence).}\label{exa:overconfidence}
The agent chooses \(a\in[0,\infty)\) to maximize \(\EE[\pi(a,Y)]\) with \(\pi(a,y)=y-c(a)\), where \(c\) is differentiable, strictly convex, satisfies \(c(0)=c'(0)=0\), and \(c'(a)\to\infty\).
Outcomes satisfy the linear–Gaussian equation
\[
Y=(\alpha^\ast+a)\theta^\ast+\varepsilon,\qquad \varepsilon\sim\mathcal N(0,1),
\]
for the true data-generating process, and
\[
Y=(\alpha+a)\theta+\varepsilon,\qquad \varepsilon\sim\mathcal N(0,1),\quad \theta\in[0,\overline\theta],
\]
for the perceived model, where \(\alpha>0\) is a fixed perceived ability (not learned), and $\overline\theta$ is sufficiently large (here, it suffices to have $\overline\theta \geq \theta^*$). We focus on overconfidence, \(\alpha>\alpha^\ast\).
These equations induce the kernels \(Q(\cdot\mid a)=\mathcal N((\alpha^\ast+a)\theta^\ast,1)\) and \(Q_\theta(\cdot\mid a)=\mathcal N((\alpha+a)\theta,1)\).
The action-dependent Kullback–Leibler divergence is
\[
K(\theta,a)=\tfrac12\Big((\alpha+a)\theta-(\alpha^\ast+a)\theta^\ast\Big)^2.
\]
For any action distribution \(\sigma\in\Delta[0,\infty)\), \(\int K(\theta,a)\,\sigma(da)\) is strictly convex in \(\theta\) and has a unique minimizer; hence we can restrict attention to degenerate beliefs (a single parameter \(\theta\)).
In particular, for pure play \(\sigma=\delta_a\) the unique KL minimizer is
\[
\theta^{\mathrm m}(\delta_a)=\theta^\ast+\theta^\ast\,\frac{\alpha^\ast-\alpha}{\alpha+a},
\]
which lies strictly below \(\theta^\ast\) under overconfidence (underestimation of the return to effort); the bias shrinks with \(a\) and vanishes as \(a\to\infty\).

Given belief \(\theta\), optimality requires $c'(a)=\theta$. 
The belief restriction requires the belief to coincide with the KL projection induced by \(a\): $\theta=\theta^{\mathrm m}(\delta_a)$.
A Berk--Nash equilibrium is therefore an action \(a\) solving
\begin{equation}\tag{3}
c'(a)=\theta^{\mathrm m}(\delta_a)
=\theta^\ast+\theta^\ast\,\frac{\alpha^\ast-\alpha}{\alpha+a}.
\end{equation}

Because \(c'(0)=0\) and \(c'(a)\to\infty\) while the right-hand side of \((3)\) is continuous, increasing, and bounded above by \(\theta^\ast\), a solution exists.
There may be multiple solutions since both sides of \((3)\) are increasing.
A simple sufficient condition for uniqueness is that \(c'\) be convex.
Under overconfidence, \(\theta^{\mathrm m}(\delta_a)<\theta^\ast\) for all \(a\), so any solution \(a\) to \((3)\) satisfies \(a<(c')^{-1}(\theta^\ast)\) (below the truth-based benchmark).
As misperception vanishes (\(\alpha\downarrow\alpha^\ast\)), \(\theta^{\mathrm m}(\delta_a)\uparrow\theta^\ast\) pointwise and the (set of) Berk--Nash solutions converges to \((c')^{-1}(\theta^\ast)\). $\blacklozenge$
\end{example}

\bigskip

\begin{example}[Trade with adverse selection]\label{exa:adverse}
A buyer posts a price $a\in\mathbb{R}_+$. Outcomes are $\mathbb{Y}=\{\square\}\,\cup\,\{v_1,\ldots,v_K\}$,
where $\square$ denotes ``no trade'' and $v_j\in\mathbb{R}$ is the realized buyer's value if trade occurs. Payoff is $0$ under no trade and $v-a$ under trade, so that trade happens at the buyer's price. Let $S$ be the seller's value and $V\in\{v_1,\ldots,v_K\}$ be the buyer's value, and let $p$ be their joint probability distribution. The seller follows the weakly dominant strategy of asking for their valuation, so we focus on the buyer. Let
\[
F_S(a):=\Pr(S\le a)
\quad\text{and}\quad
p_j(a):=\Pr\big(V=v_j\mid S\le a\big),\quad j=1,\ldots,K.
\]
The true consequence kernel puts mass $1-F_S(a)$ on $\square$ and, with probability $F_S(a)$, draws $V$ from the \emph{selected} probability mass function $p(a)=(p_1(a),\ldots,p_K(a))$.

\emph{Misspecification.} The buyer takes $F_S(a)$ as correct (e.g., learned from ask prices) but ignores selection in values—equivalently, he behaves as if $S$ and $V$ were independent, so the composition of trading sellers carries no information about $V$. Formally, he posits an \emph{action-invariant} $\theta=(\theta_1,\ldots,\theta_K)\in\Delta(\{v_1,\ldots,v_K\})$ used whenever trade occurs. Thus the perceived kernel assigns $1-F_S(a)$ to $\square$ and, with probability $F_S(a)$, chooses $V=v_j$ with probability $\theta_j$.

\emph{KL minimizer.} Fix $a$ with $F_S(a)>0$. Since true and perceived kernels agree on $\square$, the KL divergence reduces to
\[
K(a,\theta)
=F_S(a)\Big\{C(a)+\sum_{j=1}^K p_j(a)\log\!\frac{p_j(a)}{\theta_j}\Big\},
\]
with $C(a)$ independent of $\theta$. Minimizing over $\theta$ yields the \emph{unique} minimizer $\theta^{\mathrm m}(\delta_a)=p(a)$.
Consequently,
\[
\mathbb{E}_{\theta^{\mathrm m}(\delta_a)}[V]
=\sum_{j=1}^K v_j\,p_j(a)
=\mathbb{E}[V\mid S\le a].
\]

For any mixed price $\sigma\in\Delta(\mathbb{R}_+)$ with $\int F_S(a)\,\sigma(da)>0$, the KL objective averages linearly across $a$ with strictly positive weights and is strictly convex on the simplex; hence the KL minimizer $\theta^{\mathrm m}(\sigma)$ is a \emph{singleton}. Therefore, for optimality and equilibrium, we can restrict attention to \emph{degenerate} beliefs on $\theta$.

\emph{Optimal pricing under a degenerate belief.} Given $\theta$, the buyer's perceived expected payoff from posting price $a$ is
\[
U(a,\theta)=F_S(a)\big(\mathbb{E}_\theta[V]-a\big).
\]
A \emph{pure-action} Berk--Nash equilibrium with strictly positive trade is a price $a^\star$ such that $F_S(a^\star)>0$ and
\[
a^\star \in \arg\max_{a\ge 0}\; F_S(a)\Big(\EE_{\theta^{\mathrm m}(\delta_{a^\star})}[V]-a\Big).
\]
An equilibrium is thus a fixed point of the expression above reflecting the fact that the buyer forms the mean belief about $V$ using the selected distribution generated by \emph{that very price}. When contemplating deviations, he holds this belief fixed and compares only how the deviation would change the cost and the probability of trade. Under this evaluation, $a^\star$ yields at least as high a perceived payoff as any other price. $\blacklozenge$
\end{example}

\subsection{Dynamics} \label{sub:dynamics}

\paragraph{Dynamic setup.}
Fix a full-support prior \(\mu_0\in\Delta\Theta\).
For periods \(t=1,2,\ldots\):
\begin{enumerate}
\item Given the Bayesian posterior \(\mu_t\), the agent chooses \(a_t\in F(\mu_t)\).
If \(F(\mu_t)\) is multi-valued, a measurable selection \(\phi_t(\cdot\mid\mu_t)\in\Delta \mathbb{A}\) satisfies
\(\phi_t(F(\mu_t)\mid\mu_t)=1\).
\item A consequence \(y_t\) is drawn from the true kernel: \(y_t\sim Q(\cdot\mid a_t)\).
\item The posterior updates by Bayes’ rule:
\[
\mu_{t+1}(B)\;=\;\frac{\int_B q_\theta(y_t\mid a_t)\,\mu_t(d\theta)}
{\int_\Theta q_\theta(y_t\mid a_t)\,\mu_t(d\theta)},\qquad \text{$B \subseteq \Theta$ Borel}
\]
which is well-defined under the domination/LR-bound assumptions.
\end{enumerate}

\paragraph{Probability law.}
The prior \(\mu_0\), the selection rules \(\{\phi_t\}\), and the true kernel \(Q\) induce a probability measure \(\mathbf{P}\) on the infinite sequence of actions and consequences.
All almost-sure statements below are with respect to \(\mathbf{P}\). When a result depends on the prior, we write \(\mathbf{P}_{\mu_0}\) to make this dependence explicit.

\paragraph{Empirical action distribution.}
For \(t\ge1\), let
\[
\sigma_t(B)\;:=\;\frac{1}{t}\sum_{\tau=1}^t \mathbf{1}_B(a_\tau),\qquad B\subseteq\mathbb{A}\ \text{Borel}.
\]

\begin{lemma}[Posterior concentration near KL minimizers]\label{lemm:SA-concentration-open}
Almost surely, the following holds: if \(\sigma_t \Rightarrow \sigma\), then for every closed set 
\(C \subseteq \Theta \) such that \(\Theta^{\mathrm m}(\sigma) \cap C = \{\emptyset\} \), $\mu_t(C )\ \to\ 0$.
\end{lemma}

The proof of this lemma is an adaptation of Proposition \ref{pro:Berk} to a setting with endogenous learning.

\begin{proposition}\label{prop:SA-converges-to-BN}
Almost surely, the following holds: if \(a_t\to a^\star\), then \(\delta_{a^\star}\) is a Berk--Nash equilibrium; that is, there exists \(\mu^\star\in\Delta\Theta\) with \(\operatorname{supp}(\mu^\star)\subseteq \Theta^{\mathrm m}(\delta_{a^\star})\) such that $a^\star\ \in\ F(\mu^\star)$.
\end{proposition}

\begin{proof}[Sketch of the proof of Proposition~\ref{prop:SA-converges-to-BN}]
Work on the event \(\{a_t\to a^\star\}\).
Then \(\sigma_t\Rightarrow \delta_{a^\star}\). By Lemma~\ref{lemm:SA-concentration-open},
for every open \(U\supset \Theta^{\mathrm m}(\delta_{a^\star})\) we have \(\mu_t(U)\to 1\).
Hence any weak limit point \(\mu^\star\) of \((\mu_t)\) satisfies
\(\operatorname{supp}(\mu^\star)\subseteq \Theta^{\mathrm m}(\delta_{a^\star})\).
Because \(a_t\in F(\mu_t)\) and \(F\) is nonempty, compact-valued, and upper hemicontinuous,
passing to the limit yields \(a^\star\in F(\mu^\star)\).
\end{proof}

\subsection{Steady states in mixed actions (finite \(\mathbb{A}\))}

We have established that if actions converge to a pure action \(a^\star\), then \(\delta_{a^\star}\) is a Berk--Nash equilibrium. We now address \emph{mixed} steady states: distributions \(\sigma\in\Delta \mathbb{A}\) that arise as limits of mixed play.  Throughout this subsection \(\mathbb{A}\) is finite, although this can be relaxed.

\paragraph{Route I: payoff perturbations and intended mixed play (finite \(\mathbb{A}\)).}
Assume that at each \(t\), payoffs are augmented by IID action-specific shocks
\((\eta_t(a))_{a\in\mathbb{A}}\) that are independent across \(t\) and independent of histories,
with a common distribution that is absolutely continuous (with a continuous density) on \(\mathbb{R}^{|\mathbb A|}\).
This guarantees that random-utility choice probabilities are well defined and \emph{continuous}
in beliefs.

Given belief \(\mu_t\in\Delta\Theta\), define the \emph{intended mixed action} (choice probabilities)
\[
\sigma_t^{\mathrm I}(a)
\;:=\;
\mathbb{P}\!\left(
a\in\arg\max_{a'\in\mathbb{A}}
\Big\{\textstyle\int_\Theta U(a',\theta)\,\mu_t(d\theta)\;+\;\eta_t(a')\Big\}
\right),
\qquad a\in\mathbb{A}.
\]
Equivalently, there exists a continuous map \(\sigma^{\mathrm I}:\Delta\Theta\to\Delta \mathbb{A}\) such that
\(\sigma_t^{\mathrm I}=\sigma^{\mathrm I}(\mu_t)\) for all \(t\).

\begin{definition} \label{def:BeNEintended}
A distribution \(\sigma\in\Delta \mathbb{A}\) is a \emph{Berk--Nash equilibrium in intended strategies} if there exists
\(\mu\in\Delta\Theta\) such that $\operatorname{supp}(\mu)\ \subseteq\ \Theta^{\mathrm m}(\sigma)$ and $\sigma\;=\;\sigma^{\mathrm I}(\mu)$.
\end{definition}

Berk--Nash equilibria in intended strategies approximate standard Berk--Nash equilibria when payoff
perturbations are small: the smoothing inherent in random utility vanishes as noise shrinks, so intended
mixed actions approach best replies and the belief restriction remains unchanged.\footnote{Formally,
let \(\{\sigma^{\mathrm I}_{\varepsilon},\mu_{\varepsilon}\}_{\varepsilon>0}\) be Berk--Nash equilibria in intended
strategies for perturbation scale \(\varepsilon\downarrow 0\), with
\(\sigma^{\mathrm I}_{\varepsilon}=\sigma^{\mathrm I}_{\varepsilon}(\mu_{\varepsilon})\).
Under compactness of \(\mathbb A,\Theta\), continuity of \(U\), the domination/envelope conditions ensuring
continuity of \(K\), upper hemicontinuity and nonemptiness of \(\Theta^{\mathrm m}(\cdot)\), and a
vanishing-noise property \(\sigma^{\mathrm I}_{\varepsilon}(\mu)\Rightarrow \text{Prob}(F(\mu))\),
any accumulation point \((\sigma,\mu^\star)\) with \(\sigma^{\mathrm I}_{\varepsilon}\Rightarrow\sigma\),
\(\mu_{\varepsilon}\Rightarrow\mu^\star\) is a (standard) Berk--Nash equilibrium:
\(\operatorname{supp}(\sigma)\subseteq F(\mu^\star)\) and
\(\operatorname{supp}(\mu^\star)\subseteq \Theta^{\mathrm m}(\sigma)\).
The \emph{converse}—that every Berk--Nash equilibrium can be approximated by intended ones as
\(\varepsilon\to 0\)—requires additional regularity (e.g., richness of the perturbation family and
tie-handling) and is the analogue of Harsanyi’s purification theorem.}

\begin{proposition}\label{prop:steady-perturbed}
Almost surely, the following implication holds: if \(\sigma_t^{\mathrm I}\to\sigma\in\Delta \mathbb{A}\),
then \(\sigma\) is a Berk--Nash equilibrium in intended strategies.
\end{proposition}

\begin{proof}[Sketch of the proof of Proposition \ref{prop:steady-perturbed}]
(1) \emph{Intended to empirical.} Conditional on histories, \(a_t\) is drawn according to \(\sigma_t^{\mathrm I}\).
A martingale SLLN yields \(\sigma_t-\frac{1}{t}\sum_{\tau\le t}\sigma_\tau^{\mathrm I}\to0\) a.s.; hence
\(\sigma_t^{\mathrm I}\to\sigma\) implies \(\sigma_t\to\sigma\) a.s.

(2) \emph{Posterior concentration.} By Lemma~\ref{lemm:SA-concentration-open}, with probability one:
if \(\sigma_t\to\sigma\), then for every open \(U\supset\Theta^{\mathrm m}(\sigma)\) we have \(\mu_t(U)\to1\).
Thus any weak limit \(\mu^\star\) of \((\mu_t)\) satisfies \(\operatorname{supp}(\mu^\star)\subseteq \Theta^{\mathrm m}(\sigma)\).

(3) \emph{Continuity of intended choice.} Since \(\sigma^{\mathrm I}\) is continuous and
\(\sigma_t^{\mathrm I}=\sigma^{\mathrm I}(\mu_t)\to\sigma\), along any subsequence with \(\mu_t\Rightarrow\mu^\star\)
we obtain \(\sigma=\sigma^{\mathrm I}(\mu^\star)\).

Combining (2) and (3) gives the claim.
\end{proof}

\paragraph{Route II: no perturbations and steady states from empirical play (finite \(\mathbb{A}\)).}
We consider paths on which the \emph{empirical action distribution} \(\sigma_t\) converges and characterize the implied steady states.

\begin{definition} \label{def:WI} 
Weak Identification holds at $\sigma \in\Delta \mathbb{A} $ if whenever 
\(\theta,\theta'\in \Theta^{\mathrm m}(\sigma)\), we have $Q_\theta(\cdot\mid a)=Q_{\theta'}(\cdot\mid a)$ for all $a\in\operatorname{supp}(\sigma)$.
\end{definition}

Under Weak Identification, if multiple parameters tie as KL minimizers under $\sigma$, they make identical predictions for every action in $\operatorname{supp}(\sigma)$. Hence ties do not affect on-path predictions or best replies---only off-path counterfactuals may differ. Weak identification is weaker than \emph{identification}, which requires the property in Definition \ref{def:WI} to hold for \emph{all} actions, not just for those in the support of $\sigma$.

\begin{proposition}\label{prop:steady-unperturbed-WI}
Almost surely, the following implication holds: if \(\sigma_t\to \sigma\in\Delta \mathbb{A}\) and
Weak Identification holds at \(\sigma\), then \(\sigma\) is a Berk--Nash equilibrium. 
\end{proposition}
\begin{proof}[Sketch of the proof of Proposition \ref{prop:steady-unperturbed-WI}]
By Lemma~\ref{lemm:SA-concentration-open}, along \(\sigma_t\to\sigma\) we have \(\mu_t(U)\to1\) for all open \(U\supset \Theta^{\mathrm m}(\sigma)\). Any weak limit \(\mu^\infty\) thus satisfies \(\operatorname{supp}(\mu^\infty)\subseteq \Theta^{\mathrm m}(\sigma)\). 
Fix \(a\in\operatorname{supp}(\sigma)\) and take dates when \(a_t=a\); along that subsequence, optimality gives \(a\in F(\mu_t)\) and passes to the limit by upper hemicontinuity. 
Under Weak Identification, all beliefs supported on \(\Theta^{\mathrm m}(\sigma)\) induce the \emph{same} outcome distribution on \(\operatorname{supp}(\sigma)\), hence the \emph{same} expected payoffs. Therefore a single belief \(\mu^\infty\) works for every \(a\in\operatorname{supp}(\sigma)\), yielding Berk--Nash equilibrium.
\end{proof}

As the sketch of the proof suggests, without weak identification we must allow beliefs to depend on actions in equilibrium. This motivates the following generalization of equilibrium.

\begin{definition}\label{def:gBNE}
A distribution \(\sigma\in\Delta \mathbb{A}\) is a \emph{generalized Berk--Nash equilibrium} if for every \(a\in\operatorname{supp}(\sigma)\) there exists a belief \(\mu_a\in\Delta\Theta\) such that $\operatorname{supp}(\mu_a)\ \subseteq\ \Theta^{\mathrm m}(\sigma)$ and $a\ \in\ F(\mu_a)$.
\end{definition}

\begin{proposition}\label{prop:steady-unperturbed}
Almost surely, the following implication holds: if \(\sigma_t\to\sigma\in\Delta \mathbb{A}\), then \(\sigma\) is a generalized Berk--Nash equilibrium.
\end{proposition}

\medskip
\noindent\emph{Discussion.}
Route~I (perturbations) aligns with Harsanyi’s interpretation of mixed strategies and leads to Berk--Nash in intended strategies (and, under vanishing noise, to standard Berk--Nash). 
Route~II (no perturbations) views \(\sigma\) as an \emph{empirical frequency}. 
Under Weak Identification, any limit \(\sigma\) is Berk--Nash; without it, each support action can be justified by (possibly different) beliefs, giving a generalized Berk--Nash steady state.

\begin{example}[Monopolist with unknown demand]\label{exa:mono}
A monopolist chooses a price \(a\in\mathbb A=\{2,10\}\). The true mean sales at these prices is $\phi(2)=34$ and $\phi(10)=2$. Sales satisfy \(Y=\phi(a)+\varepsilon_0\) with \(\EE[\varepsilon_0\mid a]=0\) and \(Var(\varepsilon_0\mid a)<\infty\).
The monopolist’s perceived model is Gaussian with fixed variance,
$Y=\phi_\theta(a)+\varepsilon$, $\varepsilon\sim\mathcal N(0,1)$,
$\phi_\theta(a)=\alpha-\beta a$, $\theta=(\alpha,\beta)\in\Theta:=[33,40]\times[3,3.5]$.
Misspecification arises because the perfect-fit parameter for the two true means is \(\theta_0=(42,4)\notin\Theta\).

We abuse notation and let \(\sigma\) be the probability of price $10$, and $1-\sigma$ the probability of price $2$. 
With Gaussian perceived likelihood of fixed variance, the KL criterion reduces to
\begin{equation}\label{eq:K-mono}
K(\sigma,\theta)
=\tfrac12(1-\sigma)\big(34-(\alpha-2\beta)\big)^2+\tfrac12\sigma\big(2-(\alpha-10\beta)\big)^2.
\end{equation}
Given \(\theta=(\alpha,\beta)\), define perceived revenue \(R_\theta(a)=a(\alpha-\beta a)\).

First, we establish that pure prices cannot be equilibria. If \(\sigma=0\), minimizing \eqref{eq:K-mono} enforces \(\alpha-2\beta=34\), which within \(\Theta\) yields \((\alpha,\beta)=(40,3)\).
Then
\begin{align*}
R_\theta(10)-R_\theta(2)=10(\alpha-10\beta)-2(\alpha-2\beta)=8\alpha-96\beta=32>0,
\end{align*}
so \(10\) gives higher revenue (profit) than \(2\) regardless of the belief over this set of minimizers.
If \(\sigma=1\), minimizing \eqref{eq:K-mono} enforces \(\alpha-10\beta=2\), implying \(\beta\in[3.1,3.5]\) and \(\alpha=2+10\beta\in[33,37]\); then
\(8\alpha-96\beta=16-16\beta<0\), so \(2\) gives higher revenue than \(10\) regardless of the belief over the set of minimizers of \eqref{eq:K-mono}. Thus any equilibrium must strictly mix between both prices.

For interior \(0<\sigma<1\), let \(\theta^{\mathrm m}(\sigma)=(\alpha(\sigma),\beta(\sigma))\) be the (unique) minimizer of \eqref{eq:K-mono}.
Since both prices are in the support, optimality requires indifference:
\begin{equation}\label{eq:indiff}
R_{\theta^{\mathrm m}(\sigma)}(2)=R_{\theta^{\mathrm m}(\sigma)}(10)
\ \Longleftrightarrow\
2(\alpha-2\beta)=10(\alpha-10\beta)
\ \Longleftrightarrow\
\alpha=12\,\beta.
\end{equation}
For small \(\sigma\), the minimizer lies on the boundary \(\alpha(\sigma)=40\) with
\begin{equation}\label{eq:beta-of-rho}
\beta(\sigma)=\frac{3+92\sigma}{\,1+24\sigma\,}.
\end{equation}
Imposing \eqref{eq:indiff} at \(\alpha=40\) gives \(\beta=10/3\). Equating \eqref{eq:beta-of-rho} to \(10/3\) yields
\[
\frac{3+92\sigma}{1+24\sigma}=\frac{10}{3}
\ \Longrightarrow\
\sigma^\ast=\frac{1}{36}.
\]
At this \(\sigma^\ast\), $\theta^{\mathrm m}(\sigma^\ast)=\big(40,\ 10/3\big)$ and \(R_{\theta^{\mathrm m}(\sigma^\ast)}(2)=R_{\theta^{\mathrm m}(\sigma^\ast)}(10)\), so \(\sigma^\ast\) is the unique Berk--Nash equilibrium.\footnote{As \(\sigma\) increases beyond \(1/16\), the KL minimizer moves to boundary segments where \((\alpha,\beta)=(40,3.5)\) for \(\sigma\in[1/16,\,1/4]\) and \((\alpha,\beta)=(41-4\sigma,\,3.5)\) for \(\sigma\ge 1/4\). In both cases \(\alpha-12\beta<0\), implying \(R(10)-R(2)=8(\alpha-12\beta)<0\); price \(2\) strictly dominates, so no equilibrium can occur at such \(\rho\).}

Proposition \ref{prop:SA-converges-to-BN} and the fact that there is no equilibrium in pure actions imply that actions do not converge in this example. Since weak identification holds here (in fact, we have the stronger condition that, for any mixed action, there is a unique minimizer of KL divergence), by Proposition \ref{prop:steady-unperturbed-WI} we can interpret the equilibrium as a steady-state frequency of actions---the monopolist alternates between prices, but never actually mixes. Whether or not the action frequency converges to an equilibrium is a different matter, addressed later in Section \ref{sec:conv}.
\end{example}

\subsection{Forward-looking agent (non-myopic)}

We now consider a forward-looking agent with discount factor \(\delta\in(0,1)\) who chooses actions to maximize the discounted sum of expected payoffs.

\paragraph{Dynamic program.}
Let \(\bar Q_\mu(\cdot\mid a):=\int_\Theta Q_\theta(\cdot\mid a)\,\mu(d\theta)\) denote the predictive kernel under belief \(\mu\),
and let \(B(a,y,\mu)\) be the Bayesian update after observing consequence \(y\) from action \(a\).
The value function satisfies the Bellman equation
\[
W(\mu)
\;=\;
\max_{a\in\mathbb A}
\int_{\mathbb Y} \Big(\pi(a,y)+\delta\,W(B(a,y,\mu))\Big)\,\bar Q_\mu(dy\mid a),
\]
and the associated optimal-action correspondence is
\[
F_\delta(\mu)
\;:=\;
\arg\max_{a\in\mathbb A}
\int_{\mathbb Y} \Big(\pi(a,y)+\delta\,W(B(a,y,\mu))\Big)\,\bar Q_\mu(dy\mid a).
\]

\paragraph{Steady-state focus (Route II + Weak Identification).}
We consider paths on which the empirical action distribution \(\sigma_t\) converges, and we impose Weak Identification (Definition~\ref{def:WI}).

\begin{proposition}\label{prop:FL-to-BN}
Almost surely, the following implication holds: if \(\sigma_t\to\sigma\) and Weak Identification holds at \(\sigma\), then \(\sigma\) is a Berk--Nash equilibrium. 
\end{proposition}

\begin{proof}[Sketch of the proof of Proposition \ref{prop:FL-to-BN}]
By the proof of Proposition~\ref{prop:steady-unperturbed-WI} (applied to the forward-looking path with \(\sigma_t\to\sigma\)),
we obtain a \emph{single} belief \(\mu^\infty\in\Delta\Theta\) with \(\operatorname{supp}(\mu^\infty)\subseteq \Theta^{\mathrm m}(\sigma)\)
such that \(a\in F_\delta(\mu^\infty)\) for every \(a\in\operatorname{supp}(\sigma)\).
It remains to pass from \(F_\delta\) to \(F\).

By Weak Identification at \(\sigma\), $B(a,y,\mu^\infty)=\mu^\infty$, $\bar Q_{\mu^\infty}(\cdot\mid a)$-almost all $y$ and  for all $a\in\operatorname{supp}(\sigma)$. 
Therefore, for such \(a\),
\[
\int_{\mathbb Y}\!\big(\pi(a,y)+\delta W(B(a,y,\mu^\infty))\big)\,\bar Q_{\mu^\infty}(dy\mid a)
=\int_{\mathbb Y}\!\pi(a,y)\,\bar Q_{\mu^\infty}(dy\mid a)+\delta W(\mu^\infty),
\]
so the continuation term \(\delta W(\mu^\infty)\) is common across actions. Hence maximizing the dynamic
objective coincides with maximizing the myopic expected payoff on \(\operatorname{supp}(\sigma)\), i.e.,
\(F_\delta(\mu^\infty)=F(\mu^\infty)\). Since \(a\in F_\delta(\mu^\infty)\), it follows that
\(a\in F(\mu^\infty)\) for all \(a\in\operatorname{supp}(\sigma)\). Together with
\(\operatorname{supp}(\mu^\infty)\subseteq \Theta^{\mathrm m}(\sigma)\), this shows \(\sigma\) is Berk--Nash.
\end{proof}

\noindent
\emph{Interpretation.} Under Weak Identification, learning has no marginal value at the putative steady state: the next-period belief equals the current belief almost surely for support actions, so exploration does not tilt incentives. Hence forward-looking optimality coincides with myopic optimality at the limit, and the steady state is a Berk--Nash equilibrium.

\subsection{Bibliographic notes}

\cite{esponda2016berk} develop a multi-agent misspecified-learning framework that captures both systematic biases and limits to accurate environmental representation arising from complexity or informational constraints. They introduce \emph{Berk--Nash equilibrium} and show that whenever behavior converges, it converges to a Berk--Nash equilibrium. The definitions and results used here specialize their environment to a single agent and---as we show in later sections---extend straightforwardly to multiple agents. \cite{esponda2016berk} restrict the action space to be finite and establish existence under assumptions similar to those imposed here, while \citet{anderson2024existence} extend existence to compact action spaces. \cite{esponda2016berk} justify Berk--Nash equilibrium via payoff perturbations (Route I), an approach originating in \citet{fudenberg1993learning} for the convergence to mixed-strategy Nash equilibria. Route II---interpreting mixed strategies as empirical frequencies---originates in \citet{esponda2021equilibrium} and is further developed in \citet{esponda2021asymptotic}, which also highlights the need for a generalized Berk--Nash notion; this generalized concept is extended to games by \citet{murooka2023convergence}.

\cite{spiegler2016bayesian} develops the static notion of ``personal equilibrium,'' which can be read as a single-agent specialization of Berk--Nash without dynamics. The paper's distinctive contribution is to capture misspecification with directed acyclic graphs (DAGs), yielding a concise graphical language for \emph{causal} misperceptions and qualitative misunderstandings of correlation structure---especially mistaken conditional-independence assumptions and attribution errors. At the same time, the framework is not meant to capture more parametric forms of misspecification, such as incorrect functional forms. It also introduces small perturbations (trembles) to handle zero-probability events, providing a pragmatic route to identification of beliefs within the model. 

Example \ref{exa:overconfidence} (effort) is a special case of the model studied by \cite{heidhues2018unrealistic}. They show that overconfidence can make learning self-defeating: as the agent updates beliefs about an external fundamental, he becomes systematically too pessimistic, adjusts actions in the wrong direction, and ends up farther from the optimum—especially when mistakes are more costly.

Example \ref{exa:adverse} (trade) is from \cite{esponda2008behavioral}. Agents exhibit a form of correlation neglect---treating sellers’ ask prices and buyers’ valuations as independent---which induces selection neglect: they correctly perceive the average quality of traded objects but fail to recognize that higher prices would draw higher-quality offers. Consequently they post prices that are too low, exacerbating adverse selection. This idea generalizes to other adverse selection environments.

Example \ref{exa:mono} follows \citet{nyarko1991learning}, an early precursor to the literature surveyed in this chapter. He shows that when the monopolist’s model is misspecified, the price fails to converge, whereas under correct specification the price converges to the optimal monopoly level. 
\citet{esponda2016berk} use this example to show that pure-action Berk--Nash equilibria may fail to exist while a mixed-strategy equilibrium does.  Via stochastic approximation, they prove convergence to this mixed equilibrium when the monopolist learns a single demand parameter under payoff perturbations. Subsequent work uses this example to illustrate more general (non)convergence results (e.g., \cite{fudenberg2017active,frick2023belief}) and the robustness of the misspecification (e.g., \cite{fudenberg2023misspecifications}). \cite{fudenberg2017active} shows that when agents are forward-looking, they may perceive a continual benefit from experimentation, leading to non-convergent behavior even in environments where myopic agents’ actions would converge.

%% file: Vol2_ch6_game.tex
\section{Games} \label{sec:game}

For simplicity, we consider complete information games. There is a finite set of players \(I\). For each \(i\in I\), the individual action space is \(\mathbb{A}^i\); the joint action space is \(\mathbb{A}:=\times_{j\in I}\mathbb{A}^j\).
Player \(i\)'s consequences take values in \(\mathbb{Y}^i\).
The true environment is \(Q^i:\mathbb{A}\to\Delta\mathbb{Y}^i\), assigning to each action profile a distribution over \(i\)'s consequences.
Player \(i\) entertains a parametric model \(\{Q^i_{\theta^i}:\theta^i\in\Theta^i\}\), with \(Q^i_{\theta^i}:\mathbb{A}\to\Delta\mathbb{Y}^i\).
Payoffs are \(\pi^i:\mathbb{A}^i\times\mathbb{Y}^i\to\mathbb{R}\). We make the same assumptions as in Section \ref{sec:single}, with superscripts indicating a player; e.g., $\nu^i$ is the dominating measure of player $i$, etc.

A \emph{strategy} for player \(i\) is a probability distribution \(\sigma^i\in\Delta \mathbb{A}^i\).
A (mixed) \emph{strategy profile} is \((\sigma^i)_{i\in I}\in\times_{i\in I}\Delta \mathbb{A}^i\).
We restrict attention to independent mixing: the induced action-profile distribution is the product
\[
\sigma=\prod_{j\in I}\sigma^j \ \in\ \Delta \mathbb{A},
\]
and we write \(\sigma^{-i}:=\prod_{j\neq i}\sigma^j\in\Delta \mathbb{A}^{-i}\).

\paragraph{Optimal action correspondence.}
Given \(\theta^i\in\Theta^i\) and opponents’ independent mix \(\sigma^{-i}\), define
\begin{equation}\label{eq:EU-game-ind}
U^i(a^i,\sigma^{-i},\theta^i)
\;:=\;
\int_{\mathbb{A}^{-i}}\int_{\mathbb{Y}^i}
\pi^i(a^i,y^i)\,Q^i_{\theta^i}(dy^i\mid a^i,a^{-i})\,\sigma^{-i}(da^{-i}).
\end{equation}
Given a belief \(\mu^i\in\Delta\Theta^i\), the best-reply correspondence is
\[
F^i(\mu^i,\sigma^{-i})
\;:=\;
\arg\max_{a^i\in\mathbb{A}^i}
\int_{\Theta^i} U^i(a^i,\sigma^{-i},\theta^i)\,\mu^i(d\theta^i).
\]

\paragraph{KL divergence.}
For \(i\in I\), \(\theta^i\in\Theta^i\), and \(a\in\mathbb{A}\), let
\[
K^i(\theta^i,a)
\;:=\; KL(Q^{i}(\cdot \mid a) \Vert Q^{i}_{\theta^{i}}(\cdot \mid a) )
\; =\;
\int_{\mathbb{Y}^i}
\ln\!\Bigl(\frac{q^i(y^i\mid a)}{q^i_{\theta^i}(y^i\mid a)}\Bigr)\,
q^i(y^i\mid a)\,\nu^i(dy^i).
\]
For a product profile \(\sigma=\prod_{j}\sigma^j\), player \(i\)'s KL-minimizing set is
\[
\Theta^{m,i}(\sigma)
\;:=\;
\arg\min_{\theta^i\in\Theta^i}
\int_{\mathbb{A}} K^i(\theta^i,a)\,\sigma(da).
\]

\begin{definition}
A strategy profile \((\sigma^i)_{i\in I}\)  is a \emph{Berk--Nash equilibrium} if, for each \(i\in I\), there exists a belief \(\mu^i\in\Delta\Theta^i\) such that
\[
\operatorname{supp}(\mu^i)\ \subseteq\ \Theta^{m,i}(\sigma)
\qquad\text{and}\qquad
\operatorname{supp}(\sigma^i)\ \subseteq\ F^i\!\big(\mu^i,\ \sigma^{-i}\big).
\]
\end{definition}

\begin{example}{Two-player effort with interaction (Berk--Nash).}\label{exa:2.player.effort}
Players \(i\in\{1,2\}\) choose efforts \(a^i\in[0,\infty)\) and receive \(\pi^i(a^i,y^i)=y^i-c(a^i)\).
The true and perceived outcome equations are
\[
Y^i=\bigl(\alpha^\ast+a^1+a^2+a^1a^2\bigr)\theta^\ast+\varepsilon^i,
\qquad
Y^i=\bigl(\alpha+a^1+a^2+a^1a^2\bigr)\theta+\varepsilon^i,
\]
with common true ability \(\alpha^\ast>0\), perceived ability \(\alpha>\alpha^*\) (fixed overconfidence, not learned), \(\theta^\ast>0\), and \(\varepsilon^i\sim\mathcal N(0,1)\).
Assume \(c\) is continuously differentiable, strictly convex, \(c(0)=c'(0)=0\), and \(c'(a^i)\to\infty\) as \(a^i\to\infty\).\footnote{We also impose the asymptotic marginal-cost dominance
\(\displaystyle \liminf_{a^i\to\infty}\frac{c'(a^i)}{a^i}>\theta^\ast\).
This guarantees existence of a pure-strategy equilibrium when \(\theta^\ast\) is known (the truthful benchmark); the same bound delivers an invariant compact set for best responses in the Berk--Nash problem below.}

Let \(s(a):=a^1+a^2+a^1a^2\). For any (possibly mixed) action-profile distribution \(\sigma\in\Delta \mathbb{A}\), the Gaussian KL objective for either player is
\[
\int_{\mathbb{A}} \tfrac12\Big((\alpha+s(a))\theta-(\alpha^\ast+s(a))\theta^\ast\Big)^2\,\sigma(da),
\]
which is \emph{strictly convex} in \(\theta\) (unless \(\sigma\) puts mass only on profiles with \(\alpha+s(a)=0\), which cannot happen since \(\alpha>0\)).
Hence, for every \(\sigma\), the KL-minimizer \(\theta^{\mathrm m}(\sigma)\) is \emph{unique}.
In particular, for a pure profile \(a\) one has
\[
\theta^{\mathrm m}(\delta_a)=\theta^\ast\,\frac{\alpha^\ast+s(a)}{\alpha+s(a)}\ (<\theta^\ast\ \text{since }\alpha>\alpha^\ast).
\]
Therefore we may restrict attention to \emph{degenerate} beliefs (a single parameter value) in equilibrium statements.

Given a point belief \(\theta\), player \(i\)'s myopic first-order condition is $c'(a^i)=\theta\,(1+a^{-i})$.
In a Berk--Nash equilibrium (pure strategies) the belief equals the KL minimizer at the realized profile, yielding
\begin{equation}\label{eq:BN-system-2p}
c'(a^i)=(1+a^{-i})\,\theta^\ast\,\frac{\alpha^\ast+s(a)}{\alpha+s(a)},
\qquad i=1,2.
\end{equation}

For each fixed \(a^{-i}\), the right-hand side of \eqref{eq:BN-system-2p} is continuous in \(a^i\), bounded above by \((1+a^{-i})\theta^\ast\), and converges to this bound as \(a^i\to\infty\).
Because \(c'\) is continuous, strictly increasing, \(c'(0)=0\), and \(c'(a^i)\to\infty\), each equation in \eqref{eq:BN-system-2p} has a unique solution in \(a^i\) for fixed \(a^{-i}\) (single-valued best responses).
The asymptotic dominance condition \(\liminf_{a^i\to\infty} c'(a^i)/a^i>\theta^\ast\) implies that best responses map some box \([0,M]^2\) into itself, so a pure-strategy Berk--Nash equilibrium exists by Kakutani.
Equilibrium need not be unique in general (two increasing best responses can intersect more than once).
\hfill\(\blacklozenge\)
\end{example}

\subsection{Bibliographic notes}

The definition of Berk--Nash equilibrium in games follows \citet{esponda2016berk}. Above, we stated it only in two special cases: (i) complete information, and (ii) correct beliefs about the distribution of others’ play. \citet{esponda2016berk} develop the general formulation, which in particular allows for asymmetric (private) information. Allowing for such information or signals is both realistic and useful: it broadens the range of applications and facilitates identification. For example, in our earlier examples with overconfident agents, identification (uniqueness of the KL minimizer) was obtained by imposing a dogmatic belief about the degree of overconfidence. In many applications, however, one can instead obtain identification from variation in players’ actions, and such variation arises naturally when actions are conditioned on signals that may be correlated with the state of the world. The equilibrium approach then yields concrete predictions: once the relevant parameters are (at least partially) identified, Berk--Nash equilibrium allows us to characterize steady-state outcomes directly, without having to solve for the full transient learning and experimentation dynamics starting from arbitrary initial beliefs.

Example \ref{exa:2.player.effort} (effort game) is from \cite{ba2023multi}. More generally, they extend \cite{heidhues2018unrealistic}'s single‐agent model to two players who learn and best‐respond each period, so each player’s effort affects the informativeness of the other’s signal (informational externalities). With \emph{positive} externalities, one player’s higher effort raises the value of the other’s learning and induces higher effort in return---i.e., learning is \emph{mutually reinforcing}; with \emph{negative} externalities it is \emph{mutually limiting}. Because one agent's extra effort benefits the other through these externalities, a small degree of overconfidence can push efforts up and thereby improve welfare.

Results analogous to Section~\ref{sec:single} (single agent) hold for games: if play converges, it converges to a Berk--Nash equilibrium. The two routes highlighted there carry over: with independent payoff perturbations one obtains a mixed–strategy Berk--Nash equilibrium; without perturbations, one must work with a generalized equilibrium (see \cite{esponda2021asymptotic} and its extension to games in \cite{murooka2023convergence}).

The restriction that players have correct beliefs about the distribution of others' play can be rationalized if actions are publicly observed each period and agents consistently estimate the empirical distribution; for finite action sets, one can model beliefs as tracking empirical frequencies, as in (stochastic) fictitious play \citep{fudenberg1993learning}. Crucially, this introduces another misspecification: agents learn \emph{as if} others play a stationary (possibly mixed) strategy each period---generally false, though asymptotically accurate if strategies converge. This perspective illuminates Nash---and, by extension, Berk--Nash---as fixed points of mutual best responses rather than outcomes of efforts to steer others’ learning: if a player knew the opponents' learning rules, she could try to manipulate them, and the resulting steady state need not be Nash. Once players model (and try to influence) others’ learning rules, one must also model what they believe about others’ models, and what they believe others believe about \emph{their} models, and so on, generating an infinite regress. Nash and Berk--Nash close off this higher–order regress by positing best responses to the distribution of (observed) play.

While the previous discussion highlights the challenges of incorporating higher–order beliefs into (misspecified) learning, \cite{murooka2023convergence} take a concrete approach.\footnote{\cite{ba2023multi} also make common knowledge assumptions when studying how players learn in the dynamic version of their game.} They study learning through equilibrium play when outcomes follow
\[
y_t \;=\; Q\!\big(a_1^t,\,a_2^t;\,\theta^*,\,\alpha\big)\;+\;\varepsilon_t,
\]
where $\theta^*$ is scalar-valued unknown payoff–relevant state, $\alpha$ is an additional payoff parameter about which players hold \emph{dogmatic, fixed} higher–order beliefs, and $\varepsilon$ is standard Normal. Each period, given (i) their current posterior over $\theta$ and (ii) the fixed belief profile about $\alpha$, players compute the \emph{unique} Nash equilibrium of the stage game under those beliefs and play it. Afterward, they observe only the public outcome $y_t$ (not each other’s actions) and update beliefs about $\theta$ using three pieces of data: the realized $y_t$, their own action $a_i^t$, and the \emph{forecasted} opponent action implied by the equilibrium calculation (the opponent’s actual action is unobserved). Under \emph{full projection}\footnote{The paper also considers more general assumptions on (higher–order) beliefs, in the spirit of \cite{gagnon2021projection} and \cite{madarasz2023projective}.}, each player assumes others share her belief about $\theta$ and the fixed $\alpha$ at all higher orders, so the forecast equals the opponent’s equilibrium action under that belief profile. The updated posterior over $\theta$, with beliefs about $\alpha$ held fixed, feeds into the next period’s equilibrium computation, and the process repeats. Under some technical assumptions, they find that even for arbitrarily small misspecifications, there is zero probability that the posterior remains in any neighborhood of the true $\theta^*$.

%% file: Vol2_ch6_convergence.tex
\section{Convergence} \label{sec:conv}

We discuss three general approaches for determining convergence of beliefs and behavior.

\subsection{Convergence of actions} \label{sub:conv_action}

If the actions converge, we know from Proposition \ref{prop:SA-converges-to-BN} that they must converge to a pure-action Berk-Nash equilibrium. A corollary is that in a game with no pure action equilibrium, actions do not converge. This is the case, for instance, in the monopoly with unknown demand example from Section \ref{sec:single}. But even if pure action Berk--Nash equilibria exist, we still do not know if actions converge to one of these equilibria.

We restrict attention to finite action and consequence spaces and a single agent who is possibly nonmyopic.

Consider the following refinement of (pure-action) Berk-Nash equilibrium.

\begin{definition} Action \(a\) is a \emph{uniformly strict} Berk--Nash equilibrium if for every belief \(\mu\in\Delta\Theta\) with \(\operatorname{supp}(\mu)\subseteq\Theta^{\mathrm m}(\delta_a)\) we have \(F(\mu)=\{a\}\).
\end{definition}

While a converse to Proposition \ref{prop:SA-converges-to-BN} does not hold, it turns out that if we strengthen the equilibrium to be uniformly strict, then we obtain a necessary and sufficient condition for the following notion of stability:

\begin{definition}
An action \(a\in\mathbb A\) is \emph{uniformly stable} if for every \(\kappa\in(0,1)\) there exists \(\varepsilon>0\) such that, for every prior \(\mu_0\in\Delta\Theta\) with \(\mu_0\!\left(\widehat{\Theta}^\varepsilon(a)\right)>1-\varepsilon\), the action prescribed by any optimal policy converges to \(a\) with probability at least \(1-\kappa\), where
\[
\widehat{\Theta}^\varepsilon(a)
:=\Bigl\{\theta\in\Theta:\ \exists\,\tilde\theta\in\Theta^{\mathrm m}(\delta_a)\ \text{such that}\  || Q_\theta(\cdot\mid a) - Q_{\tilde\theta}(\cdot\mid a) || \le\varepsilon\Bigr\}.
\]
\end{definition}

Roughly speaking, an action is uniformly stable if play converges to it with high probability for any optimal policy and starting from every belief in a neighborhood of a KL minimizer for that action.

\begin{proposition}
An action $a$ is uniformly stable if and only if it is a uniformly strict Berk--Nash equilibrium.
\end{proposition}

The idea behind the ``if" part is a self-reinforcing loop: if \(a\) is uniformly strict, there is a neighborhood \(B\) of its KL-minimizers and a margin \(m>0\) such that whenever the posterior places enough mass on \(B\), the myopic advantage of \(a\) is at least \(m\), while the forward-looking (experimentation) value of deviating is uniformly bounded by some \(m/2\); thus every optimal policy---myopic or nonmyopic---selects \(a\) in that region. Playing \(a\) then produces data that, with high probability, keeps the posterior inside (or quickly returns it to) \(B\), so the condition that makes \(a\) optimal tends to persist. In particular, the chance of repeatedly being pushed far enough out of \(B\) to overturn the strict margin is small and \emph{transient}; after at most finitely many such \emph{excursions}, beliefs remain in \(B\) and optimal policies keep selecting \(a\), delivering uniform stability.

We now ask whether there is a positive chance that play eventually settles at some action $a$ under any \emph{full–support} prior. 

\begin{definition}
An action \(a\in\mathbb A\) is \emph{positively attractive} if for every initial belief \(\mu_0\in\Delta\Theta\) with \(\operatorname{supp}(\mu_0)=\Theta\), $\mathbf P_{\mu_0}\!\left(\lim_{t\to\infty} a_t = a\right)\;>\;0$.
\end{definition}

One way to obtain positive attractiveness is to identify environments in which, starting from a full–support prior, there is \emph{positive probability} that beliefs enter the KL–neighborhood where a uniformly strict action \(a\) is uniquely optimal. By the uniform–stability result above, once beliefs enter that neighborhood, optimal play sticks to \(a\), so the path converges to \(a\) on that event.

\begin{definition}
Outcomes are \emph{subjectively exogenous} if, for every \(a,a'\in\mathbb A\) and every \(\theta\in\Theta\), $Q_\theta(\cdot\mid a)\;=\;Q_\theta(\cdot\mid a')$.
That is, under the agent’s model the consequence distribution does not depend on the action.
\end{definition}

\begin{proposition}
Suppose outcomes are subjectively exogenous. If \(a\) is a uniformly strict Berk--Nash equilibrium, then \(a\) is positively attractive.
\end{proposition}

When outcomes are subjectively exogenous, posterior updating depends only on realized outcomes, not on which actions produced them. Because the true outcome laws across actions are mutually absolutely continuous, there is \emph{positive probability} of observing a finite sequence of outcomes whose empirical distribution is arbitrarily close to the true distribution under action \(a\). Along any such sequence, one can show that the posterior places \emph{arbitrarily high mass} on the KL–minimizer set \(\Theta^{\mathrm m}(\delta_a)\). Since \(a\) is \emph{uniformly strict}, once beliefs are that concentrated, every optimal (including forward-looking) policy uniquely selects \(a\); from there, the earlier uniform-stability argument applies: playing \(a\) keeps beliefs in that region with high probability, so with \emph{positive probability} the process remains at \(a\) thereafter, i.e., convergence to \(a\) occurs with positive probability.


\subsection{Convergence of empirical action distribution} \label{sub:conv_distrib}

We continue to work with a finite action space, but no longer impose finiteness of the consequence space. We assume the agent is myopic (some of this can be relaxed under weak identification). Our objective is to characterize the asymptotic behavior of \emph{action frequencies}. Even when the action sequence does not converge, empirical frequencies may converge, delivering informative predictions about long-run behavior and payoffs. 

We can express the empirical action frequency \( \sigma_t \in \Delta \mathbb{A} \) recursively as
\[
\sigma_{t+1}
=\sigma_t+\frac{1}{t+1}\big(\mathbf{1}(a_{t+1})-\sigma_t\big),
\tag{5}
\]
where \(\mathbf{1}(a_{t+1})\) is the action-indicator vector. Adding and subtracting the conditional expectation given current information,
\[
\sigma_{t+1}
=\sigma_t+\frac{1}{t+1}\big(\mathbb{E}[\mathbf{1}(a_{t+1})\mid \mu_{t+1}]-\sigma_t\big)
+\frac{1}{t+1}\underbrace{\big(\mathbf{1}(a_{t+1})-\mathbb{E}[\mathbf{1}(a_{t+1})\mid \mu_{t+1}]\big)}_{\text{martingale difference}},
\]
so the evolution depends on beliefs \(\mu_{t+1}\). The last term is a martingale-difference noise with respect to the natural filtration, uniformly bounded and multiplied by a step size $1/(t{+}1)$; hence it is asymptotically negligible.

The key step is to re-express this dependence using a characterization of \emph{limit beliefs} in the spirit of Berk's result and Lemma~\ref{lemm:SA-concentration-open} in Section \ref{sec:single}: for large \(t\), \(\mu_{t+1}\) concentrates on parameter values whose KL-divergence is nearly minimal for the data generated at the current frequency \(\sigma_t\).

To handle potential discontinuities of the map \(\mu \mapsto \mathbb{E}[\mathbf{1}(a_{t+1})\mid \mu]\), we replace it by a correspondence \(\Delta F(\mu)\subseteq \Delta \mathbb{A}\) that contains this function (i.e., whose graph contains the graph of the function) and is well-behaved (upper hemicontinuous with nonempty, compact values). Then we can write the stochastic \emph{difference inclusion}
\[
\sigma_{t+1}
=\sigma_t+\frac{1}{t+1}(r_{t+1}-\sigma_t),
\qquad r_{t+1}\in \Delta F(\mu_{t+1}).
\]
Using the large-\(t\) concentration of \(\mu_{t+1}\) around the KL minimizers determined by \(\sigma_t\), the equation above can be rewritten entirely in terms of \(\sigma_t\) plus approximation errors that vanish asymptotically.

Next, view the discrete sequence \((\sigma_t)\) as points in \(\Delta \mathbb{A}\) and connect consecutive points by straight line segments, traversed on the “slow” time scale \(\tau_t=\sum_{i=1}^t 1/i\) (so steps shrink like \(1/(t+1)\)). This yields a piecewise-linear interpolation \(w(\cdot)\) that tracks \(\sigma_t\) closely. 
After absorbing the vanishing error terms, it turns out that \(w\) is well-approximated by solutions to the differential inclusion
\[
\dot{\sigma}(t)\in \Delta F(\Delta\Theta^{\mathrm m}(\sigma(t))) - \sigma(t),
\tag{DI}
\]
which serves as the continuous-time dynamical system guiding the long-run behavior of the discrete process \((\sigma_t)\).

This is stated formally in the following proposition.

\begin{proposition}\label{prop:cts-time-approx}
Let \(w:\mathbb{R}_+\to \Delta\mathbb{A}\) be the continuous-time interpolation of the empirical action frequencies \((\sigma_t)_{t\ge1}\), and consider the differential inclusion (DI). 
For an initial condition \(\sigma(0)\in \Delta\mathbb{A}\), let \(\mathcal{S}_{\sigma(0)}\) denote the set of absolutely continuous solutions \(\sigma(\cdot)\) to \textup{(DI)} with that initial condition, and let \(\mathcal{S}\equiv \bigcup_{\sigma(0)\in\Delta\mathbb{A}}\mathcal{S}_{\sigma(0)}\) be the set of all such solutions.

Then, almost surely, for every \(T>0\),
\[
\lim_{t\to\infty}\;
\inf_{\;\sigma(\cdot)\in \mathcal{S}}\;
\sup_{s\in[0,T]}
\big\|\,w(t+s)\;-\;\sigma(s)\,\big\| \;=\; 0.
\]
\end{proposition}

To understand this statement, fix any finite horizon length \(T>0\). Look far enough into the process (take \(t\) large). Then there exists a solution \(\sigma^{t}(\cdot)\) of the differential inclusion (DI) that
\emph{starts arbitrarily close} to the current discrete state \(w(t)\) and whose
trajectory \emph{shadows} the interpolated path thereafter on the time window \([0,T]\).
Formally: for every \(\varepsilon>0\) and \(T>0\) there is \(t_0\) such that for all \(t\ge t_0\) one can find a solution \(\sigma^{t}(\cdot)\) to (DI) with
\[
\|\sigma^{t}(0)-w(t)\|\le \varepsilon
\quad\text{and}\quad
\sup_{s\in[0,T]}\big\|\sigma^{t}(s)-w(t+s)\big\|\le \varepsilon.
\]
We do \emph{not} insist that the DI solution pass exactly through \(w(t)\) at \(s=0\); small stochastic and discretization errors generally prevent exact alignment. Instead, the statement guarantees that, asymptotically, there is a DI solution starting \emph{arbitrarily close} to \(w(t)\) that remains uniformly close to the interpolated discrete path over any fixed finite window. 

A rest point of the DI is a generalized Berk-Nash equilibrium (see Definition \ref{def:gBNE}): for every $a\in\operatorname{supp}(\sigma)$ there exists $\mu_a\in\Delta\Theta(\sigma)$ such that $a\in F(\mu_a)$.
Unlike standard Berk--Nash (which uses a \emph{single} belief to justify all support actions), the generalized version allows different support actions to be justified by \emph{different} beliefs, each drawn from the KL–minimizing set consistent with $\sigma$.
This generalized notion is natural when the object of interest is an \emph{action distribution}: if beliefs do not settle down, distinct support actions may be chosen under genuinely different (yet $\sigma$-consistent) beliefs.
Analogous to the results in Section~\ref{sec:single}, one can show that if $\sigma_t \to \sigma$, then the limit $\sigma$ is a generalized Berk--Nash equilibrium.

Next, we turn to the question of convergence. Define $d(\sigma,\Sigma) := \inf_{\tilde{\sigma}\in \Sigma} \|\sigma - \tilde{\sigma}\|$.

\begin{definition}
A set $\Sigma\subseteq \Delta\mathbb{A}$ is \emph{attracting} if there exists an open $U$ with $\Sigma\subset \operatorname{int}U$ such that for every $\varepsilon>0$ there is $T>0$ with
$d(\sigma(t),\Sigma)<\varepsilon$ for all $t>T$ and for every solution $\sigma(\cdot)$ of \textup{(DI)} with initial condition $\sigma(0)\in U$.
The largest such $U$ is the \emph{basin of attraction} $U_\Sigma$.
The set $\Sigma$ is \emph{globally attracting} if $U_\Sigma=\Delta\mathbb{A}$.
\end{definition}

Globally attracting means uniform pull: for every $\varepsilon>0$ there exists a time $T(\varepsilon)$ such that, simultaneously for \emph{all} initial conditions $\sigma(0)\in\Delta\mathbb A$ and \emph{all} solutions $\sigma(\cdot)$ of the DI starting there, we have $d(\sigma(t),\Sigma)<\varepsilon$ for every $t\ge T(\varepsilon)$. Thus a single clock works for everyone---one does not choose $T$ solution-by-solution or initial-condition-by-initial-condition. This is stronger than merely saying that each solution converges to $\Sigma$, which allows $T$ (or rates) to depend on the particular trajectory; global attraction requires uniform eventual closeness over the entire state space.

\begin{proposition}\label{prop:global-attraction}
If $\Sigma\subseteq \Delta\mathbb{A}$ is globally attracting, then $\lim_{t\to\infty} d(\sigma_t,\Sigma)=0$ almost surely. In particular, if $\Sigma=\{\sigma^\ast\}$ is a globally attracting equilibrium, then $\sigma_t\to\sigma^\ast$ a.s.
\end{proposition}

The key in the proof of Proposition \ref{prop:global-attraction} is to compare the discrete path to DI solutions on fixed-length windows and then chain these comparisons. Global attraction gives a \emph{single} time $T(\varepsilon)$ that works for \emph{all} DI solutions, regardless of where they start. Proposition~\ref{prop:cts-time-approx} lets us pick, for large $t$, a DI solution that stays within $\varepsilon$ of the interpolated path $w(t+\cdot)$ on $[0,T]$. Because \emph{every} DI solution is within $\varepsilon$ of $\Sigma$ after time $T$, the comparison solution is $\varepsilon$-close to $\Sigma$ at $T$, hence $w(t+T)$ is within $2\varepsilon$ of $\Sigma$, and (unwinding the interpolation) $\sigma_{t'}$ is within $3\varepsilon$ of $\Sigma$ for nearby $t'$. Repeating this on successive windows, the uniform pull of global attraction prevents the process from escaping the $\varepsilon$-neighborhood of $\Sigma$, and since $\varepsilon>0$ is arbitrary, $d(\sigma_t,\Sigma)\to 0$ almost surely.

\begin{example}
We now revisit the monopolist with unknown demand example from Section \ref{sec:single}. 
Recall the example has a unique Berk--Nash equilibrium, given by mixing probability $\sigma^\ast=\tfrac{1}{36}$ on price $10$.
Writing $\sigma(t)$, the DI collapses to the one-dimensional inclusion
\[
\dot\sigma(t)\ \in\ 
\begin{cases}
\{\,1-\sigma(t)\,\}, & \sigma(t)<\sigma^\ast,\\[2pt]
[-\sigma(t),\,1-\sigma(t)], & \sigma(t)=\sigma^\ast,\\[2pt]
\{\,-\sigma(t)\,\}, & \sigma(t)>\sigma^\ast,
\end{cases}
\tag{DI-mono}
\]
which simply says: the mass on price $10$ moves toward the current best reply—rising at rate $1-\sigma$ when $10$ is uniquely optimal, falling at rate $\sigma$ when $2$ is uniquely optimal, and (under indifference) evolving with any velocity between these two extremes.

Define \(I_\varepsilon = [L, U]\) with 
\(L = \max\{0, \sigma^* - \varepsilon\}\) and \(U = \min\{1, \sigma^* + \varepsilon\}\). 
To show that \(\{\sigma^*\}\) is globally attracting, we verify that all trajectories 
enter and remain within \(I_\varepsilon\) after some finite time \(T(\varepsilon)\) 
that does not depend on the initial condition. 
At the boundaries, the vector field always points inward: 
if \(L = 0\), then \(\dot\sigma = 1 > 0\); if \(L = \sigma^* - \varepsilon > 0\), 
then \(\dot\sigma = 1 - \sigma \ge 1 - (\sigma^* + \varepsilon) > 0\); 
if \(U = 1\), then \(\dot\sigma = -1 < 0\); and if \(U = \sigma^* + \varepsilon < 1\), 
then \(\dot\sigma = -\sigma \le -(\sigma^* + \varepsilon) < 0\). 
Hence, any trajectory that enters \(I_\varepsilon\) remains there. 
If \(\sigma(0) \le \sigma^* - \varepsilon\) (so $\varepsilon \leq 1/36$), then while \(\sigma(t) < \sigma^*\), 
we have \(\dot\sigma(t) = 1 - \sigma(t) \ge 1 - \sigma^* = 35/36\), implying that 
the time to reach \(I_\varepsilon\) satisfies 
\[
T_\downarrow(\varepsilon) \le \frac{(\sigma^* - \varepsilon) - \sigma(0)}{35/36} 
\le \frac{36}{35}\Big(\tfrac{1}{36} - \varepsilon\Big) \le \tfrac{1}{35}.
\]
If \(\sigma(0) \ge \sigma^* + \varepsilon\) (so $\varepsilon \leq 35/36$), then while \(\sigma(t) > \sigma^*\), 
we have \(\dot\sigma(t) = -\sigma(t) \le -\sigma^* = -1/36\), so 
\[
T_\uparrow(\varepsilon) \le 
\frac{\sigma(0) - (\sigma^* + \varepsilon)}{1/36} 
\le 35 - 36\varepsilon \le 35.
\]
Thus, every trajectory enters \(I_\varepsilon\) within $T(\varepsilon) \le \max\{1/35,\,35 - 36\varepsilon\} \le 35$ and remains there forever. Therefore, for all \(t \ge T(\varepsilon)\), 
\(|\sigma(t) - \sigma^*| \le \varepsilon\), and by definition 
\(\{\sigma^*\}\) is globally attracting. Hence, by Proposition~\ref{prop:global-attraction}, the empirical action frequencies converge almost surely to the unique mixed equilibrium $\sigma^*=1/36$. 
\end{example}

Proposition~\ref{prop:global-attraction} yields convergence results for attracting sets.
What it does \emph{not} yield is a \emph{non}-convergence result for unstable equilibria.
The continuous-time approximation guarantees that, on any fixed horizon, the interpolated path can be shadowed by \emph{some} DI solution whose initial condition is only required to be \emph{close} to the current state. 
That ``close" neighborhood may include the unstable equilibrium itself, which admits a DI solution that stays put. 
Hence, we cannot preclude convergence of the discrete process to an unstable equilibrium using this result alone.

\subsection{Convergence of beliefs} \label{sub:conv_belief}

So far we have focused on convergence of actions or action distribution. We now consider a complementary approach, convergence of beliefs. One message from Section \ref{sec:exo} is that a standard application of martingale convergence theorems is not useful here, but indeed it turns out that a more sophisticated argument can be used to establish belief convergence that relies on martingale arguments.

Previously, we studied convergence with finite action space. We now consider a continuum action space and further restrict attention to cases where there is a continuous selection from the optimal action correspondence $F$. This typically rules out finite action spaces. A sufficient condition is that for every belief there is a unique optimal action. We denote the continuous selection from $F$ by $\mu \mapsto a^*(\mu)$.

\begin{definition}\label{def:stability}
Consider any $\theta \in \Theta$. The degenerate belief $\delta_{\theta}$ is:
\begin{enumerate}
    \item \emph{locally stable} if for any $\gamma<1$, there exists a neighborhood $\mathcal{B}\ni\delta_{\theta}$ such that 
    $\mathbf{P}_{\mu}\!\left[\mu_t \to \delta_{\omega}\right]\ge \gamma$ for each initial belief $\mu\in\mathcal{B}$;
    \item \emph{unstable} if there exists a neighborhood $\mathcal{B}\ni\delta_{\omega}$ such that 
    $\mathbf{P}_{\mu}\!\left[\exists\, t:\ \mu_t \notin \mathcal{B}\right]=1$ for each initial belief $\mu\in\mathcal{B}$.
\end{enumerate}
\end{definition}

\begin{definition}\label{def:KLorder}
Fix an action $a$. For $\theta,\theta'\in\Theta$, define $\theta \succsim^{\mathrm{KL}}_{a} \theta' 
\Longleftrightarrow 
K(\theta,a)\le K(\theta',a)$.
\end{definition}

We refer to a Berk-Nash equilibrium belief as a belief that supports a Berk-Nash equilibrium. We restrict attention to degenerate beliefs.

\begin{proposition}\label{prop:BN_point}
$\delta_\theta$ is a Berk--Nash equilibrium belief if and only if $\theta \succsim^{\mathrm{KL}}_{a^*(\delta_\theta)} \theta'$ for all $\theta'\in\Theta$; 
equivalently, $\theta\in\arg\min_{\tilde\theta\in\Theta} K(\tilde\theta,a^*(\delta_\theta))$.
\end{proposition}

The following proposition is the analogous to the results in Section \ref{sec:single} that if behavior converges, it converges to a Berk-Nash equilibrium.

\begin{proposition}\label{prop:unstable_nonBN}
If $\delta_\theta$ is not a Berk--Nash equilibrium belief, then $\delta_\theta$ is \emph{unstable}.
\end{proposition}

The preorder $\succsim^{\mathrm{KL}}_{a}$ is useful for the characterization in Proposition~\ref{prop:unstable_nonBN}, but as the next example illustrates, it may not be helpful for obtaining \emph{local} stability results.

\begin{example}[Slow learning]\label{ex:costly_info_two_state}
A single agent faces a fixed but unknown true state $\theta^\ast\in\Theta=\{\theta_1,\theta_2\}$, with $\theta_2>\theta_1>0$. Each period she chooses a precision/effort level $a\in[0,\bar\gamma]$ at cost $C(a)$, then observes $y\in\{0,1\}$ with true probability $Q(1 \mid a)=a\theta^*+\beta^*$. The agent underestimates the base rate, taking $\beta<\beta^\ast$, which is a form of “ego-biased” updating: she overreacts to good news about her ability and underreacts to bad news; her perceived signal model is $Q_\theta(1\mid a)=a\theta+\beta$.  Let the myopically optimal effort be unique for each belief and represented by the continuous function $\mu \mapsto a^*(\mu)$. Assume $a^\ast(\mu)>0$ for every nondegenerate belief $\mu$ (reflecting a positive value of even small effort when the true parameter is uncertain), and $a^\ast(\mu)=0$ for any degenerate belief $\mu$. (This assumption can be obtained from natural assumptions on the primitives.)

If $a^\ast(\mu)=0$, then $z$ is uninformative (independent of $\theta$); hence $K(\theta_1,a^\ast(\mu))=K(\theta_2,a^\ast(\mu))$. By Proposition~\ref{prop:BN_point}, both point beliefs $\delta_{\theta_1}$ and $\delta_{\theta_2}$ are Berk--Nash equilibrium beliefs. Thus the KL preorder $\succsim_a^{\mathrm{KL}}$ does not discriminate locally here and is unhelpful for establishing convergence.
\end{example}

\begin{definition}\label{def:qdom}
Fix an action $a$ and $q>0$. For $\theta,\theta'\in\Theta$, define
\[
\theta \succsim^{q}_{a} \theta'
\quad\Longleftrightarrow\quad
\int \!\!\left(\frac{q_{\theta'}(y \mid a)}{q_{\theta}(y \mid a)}\right)^{\!q}\,  Q(dy\mid a)\;\le\;1.
\]
\end{definition}

The KL preorder $\succsim^{\mathrm{KL}}_{a}$ is complete, whereas $\succsim^{q}_{a}$ is generally incomplete. The family $\{\succsim^{q}_{a}\}_{q>0}$ is nested and approximates KL-dominance as $q\downarrow 0$. Moreover, if $\theta \succ^{\mathrm{KL}}_{a} \theta'$, then there exists $q>0$ such that $\theta \succ^{q}_{a} \theta'$.

\begin{proposition}\label{prop:qdom_stability}
Consider any $\theta\in\Theta$. The point belief $\delta_\theta$ is:
\begin{enumerate}
    \item \emph{locally stable} if there exist $q>0$ and a neighborhood $\mathcal{B}\ni\delta_\theta$ such that, for every $\mu\in\mathcal{B}\setminus\{\delta_\theta\}$, every action $a$ with $a=a^\ast(\mu)$, and every $\theta'\neq\theta$, we have $\theta \succ^{q}_{a} \theta'$.
    \item \emph{unstable} if there exist $q>0$, a neighborhood $\mathcal{B}\ni\delta_\theta$, and some $\theta'\neq\theta$ such that
    for every $\mu\in\mathcal{B}\setminus\{\delta_\theta\}$ and every action $a$ with $a=a^\ast(\mu)$, $\theta' \succ^{q}_{a} \theta$.
\end{enumerate}
\end{proposition}

\begin{example}[continued]
Fix \(\theta^\ast = \theta_1\). Since $Q(1\mid a) = a\theta^\ast + \beta^\ast$ and $Q_{\theta}(1\mid a) = a\theta + \beta$, we have
$| Q(1\mid a) - Q_{\theta_1}(1\mid a) | = | \beta^\ast - \beta |$ and $| Q(1\mid a) - Q_{\theta_2}(1\mid a) | = | (\beta^\ast - \beta) - a(\theta_2 - \theta_1) |$.
Because \(\theta_2 > \theta_1\) and \(\beta^\ast > \beta\), define $\bar a := \frac{\beta^\ast - \beta}{\theta_2 - \theta_1} > 0$, so that the term inside the second absolute value is positive for \(a \in (0, \bar a]\) and negative for \(a > \bar a\).
Hence, for \(a \in (0, \bar a]\),
\[
\big| Q(1\mid a) - Q_{\theta_2}(1\mid a) \big|
= (\beta^\ast - \beta) - a(\theta_2 - \theta_1)
< \beta^\ast - \beta
= \big| Q(1\mid a) - Q_{\theta_1}(1\mid a) \big|.
\]
It follows that \(K(\theta_2,a) < K(\theta_1,a)\) for all \(a \in (0, \bar a]\).
Because \(a^\ast(\delta_{\theta_1}) = 0\), \(a^\ast(\mu) > 0\) for nondegenerate \(\mu\), and \(a^\ast\) is continuous, there exists a neighborhood \(\mathcal{B}\) of \(\delta_{\theta_1}\) such that \(0 < a^\ast(\mu) \le \bar a\) for all \(\mu \in \mathcal{B} \setminus \{\delta_{\theta_1}\}\).
Hence,
\[
K\!\big(\theta_2,a^\ast(\mu)\big) < K\!\big(\theta_1,a^\ast(\mu)\big)
\quad\Longrightarrow\quad
\theta_2 \succ^{\mathrm{KL}}_{a^\ast(\mu)} \theta_1,
\]
so by (KL \(\Rightarrow q\)) there exists \(q>0\) such that 
\(\theta_2 \succ^{q}_{a^\ast(\mu)} \theta_1\) for all 
\(\mu \in \mathcal{B}\setminus\{\delta_{\theta_1}\}\).
Proposition~\ref{prop:qdom_stability} implies that 
\(\delta_{\theta_1}\) is unstable, while by a symmetric argument around \(\delta_{\theta_2}\),
\(\delta_{\theta_2}\) is locally stable.
\end{example}

\subsection{Bibliographic notes}

The material in Section~\ref{sub:conv_action} (\emph{action convergence}) is entirely due to \cite{fudenberg2021limit}. They also introduce the concept of a \emph{uniform Berk--Nash equilibrium action}, which 
coincides with Berk--Nash equilibrium under weak identification (Definition~\ref{def:WI}) but refines the 
concept when identification fails. They further strengthen Proposition~\ref{prop:SA-converges-to-BN} by showing that if the 
sequence of actions converges, it must converge to a uniform Berk--Nash equilibrium. Consequently, 
if a uniform Berk--Nash equilibrium does not exist, actions do not converge.  \cite{fudenberg2021limit} also extend their results to environments with a pre-action signal that can affect both the realized consequence and the payoff. Moreover, lack of pathwise convergence need not preclude frequency convergence: in the limit, the empirical distribution can place all mass on \(a^*\) even though play never settles there, because another action is taken infinitely often with vanishing frequency. \cite{fudenberg2021limit} give a condition that rules out this phenomenon.

The material in Section~\ref{sub:conv_distrib} (\emph{action-distribution convergence}) follows entirely from \cite{esponda2021asymptotic}. Proposition~\ref{prop:cts-time-approx} builds, among other things, on stochastic-approximation results in \cite{benaim2005stochastic}. Solving the differential inclusion can be difficult when the action set is large; \cite{esponda2021asymptotic} show that under suitable conditions one can instead work with a differential inclusion on the parameter space $\Theta$. When $\Theta$ is one-dimensional, this often simplifies the analysis and scales to many actions, though it becomes challenging again if $\Theta$ has  higher dimensions. \cite{murooka2023convergence} extend these results to models with a continuum of actions. As noted earlier, a limitation of the differential-inclusion approach is ruling out convergence to \emph{unstable} equilibria: \cite{pemantle1990nonconvergence} already emphasized this and introduced noise to obtain non-convergence (see also \cite{Benaim1999DSA}). More recently, \cite{murooka2025bayesian} give conditions under which a stochastic process converges with \emph{zero} probability to linearly unstable steady states; importantly, their result covers Gaussian noise, which is standard in applications, while earlier results do not.

The results in Section~\ref{sub:conv_belief} (\emph{belief convergence}) are entirely due to \cite{frick2023belief}. While we assumed for simplicity that actions choices are continuous in $\mu$, they establish the results under the weaker assumption that action choices are continuous in $\mu$ near point-mass beliefs, but can be discontinuous at other beliefs. As such, this condition can also be compatible with finite action spaces. In Section~\ref{sec:exo} we noted that the martingale convergence theorem yields only \emph{subjective} belief convergence and thus is uninformative under misspecification. \cite{frick2023belief} restore a martingale-style argument by shifting the object of analysis. Their key observation is that, on any belief region where the $q$-dominance order holds, the $q$-th power of the posterior ratio becomes a nonnegative supermartingale. This insight allows them to import martingale convergence techniques from the correctly specified case and serves as the central tool for their results.

We presented their ideas in an action-choice setting; \cite{frick2023belief} formulate them in a more general, abstract environment without explicit agents, actions, or preferences. This framework naturally covers certain misspecified social-learning environments. While \cite{frick2023belief} assume a finite parameter space, they show that some of their results extend to to compact, finite-dimensional spaces. 

Beyond local stability, \cite{frick2023belief} also prove a global result: 
beliefs converge to the set that survives the iterated elimination of KL-dominated beliefs. This result requires continuity of action choices in \(\mu\) for all beliefs, not only near point-mass beliefs.
In the slow-learning example, their analysis extends to more than two parameter values and delivers global convergence to the highest parameter value, regardless of the truth. In particular, they show that vanishingly small amounts of misspecification can lead to extreme failures of (correct) learning.

Taken together, the three convergence approaches cover a broad set of applications but are complementary and best suited to different settings. The \emph{action-convergence} approach assumes a finite action and consequence sets and does not handle cases where actions do not converge even though their long-run frequencies do. The \emph{empirical-distribution} approach characterizes the asymptotic distribution of actions, whether or not actions converge. However, it requires solving a differential inclusion, which can be challenging in high-dimensional action or parameter spaces. The \emph{belief-convergence} approach accommodates a continuum of actions, though several applications rely on finiteness of the parameter space, which can be restrictive. In addition, it requires that actions are continuous in beliefs near point-mass beliefs---or everywhere, in the case of global stability results.

For applied work, a combination of these results---or a modest adaptation---will often deliver the needed (non-)convergence conclusions, though some technical effort may be required. More recently, \citet{EspondaPouzo2025-BNR} introduce \emph{Berk--Nash rationalizability}, defined in direct analogy to classical rationalizability: (i) there is an operator $\Gamma$, analogous to the best–response operator, and (ii) a set of actions $A$ is \emph{self-justified} if $A \subseteq \Gamma(A)$. For example, in the single-agent case,
\[
\Gamma(A)
= \bigl\{\, a \in \mathbb{A} \;:\; \exists \,\sigma \in \Delta A,\ \mu \in \Delta \Theta^{m}(\sigma)\ \text{such that}\ a \in F(\mu) \,\bigr\}.
\]
The set of rationalizable actions is the largest self-justified set and can be obtained by iterating $\Gamma$ from the full action set $\mathbb{A}$. They show that, almost surely, every action that is played (or approached) infinitely often is rationalizable. In this sense, it is the action-space analogue of \cite{frick2023belief}’s iterative characterization of globally stable beliefs; crucially, it does \emph{not} require a continuous selection of $F$ and therefore applies to finite action sets and to environments with multiple optimal actions for a given belief. When the Berk--Nash rationalizable set is a singleton, this yields almost-sure convergence to that action.

Beyond these general approaches, there is a rich line of results that establish (non-) convergence in specific misspecified settings. \citet{nyarko1991learning} shows in the monopoly-with-unknown-demand example of Section~\ref{sec:single} that actions do not converge. \citet{schwartzstein2014selective} proves convergence of beliefs in a selective-attention model. \citet{fudenberg2017active} analyze a continuous-time learning model with Brownian noise (without assuming myopia) and provide conditions for action convergence. They also show that forward-looking agents may perceive a continual benefit from experimentation, 
causing their actions to fail to converge in situations where a myopic agent’s actions would converge. \citet{heidhues2018unrealistic} derive global convergence to a unique Berk--Nash equilibrium by a nonmyopic agent using a contraction mapping argument in additively separable, supermodular environments. \citet{he2022mislearning} adapts this approach to a censored-data, gambler’s-fallacy setting and \cite{ba2023multi} extend the contraction-based method to multi-agent learning; an example is the effort game in Section~\ref{sec:game}. For two-point priors, \citet{bohren2021learning} give necessary and sufficient conditions for action/belief convergence and provides general local and global stability criteria based on the Kullback--Leibler divergence order.
Finally, \citet{heidhues2021convergence} prove convergence to a Berk--Nash equilibrium under normal priors and normal signals.

There is also a complementary literature on misspecified social learning. \citet{bohren2016informational} studies a setting where agents misperceive the share of informed predecessors and show convergence to steady states; \citet{frick2020misinterpreting} analyze environments in which agents misperceive the population’s type distribution and prove convergence. They also highlight the fragility of (correct) social learning under even tiny misspecifications.

%% file: Vol2_ch6_references.bib
@article{Schwartz1965,
  author  = {Schwartz, Lorraine},
  title   = {On Bayes procedures},
  journal = {Zeitschrift f{\"u}r Wahrscheinlichkeitstheorie und Verwandte Gebiete},
  year    = {1965},
  volume  = {4},
  pages   = {10--26}
}

@book{evans2001learning,
  title={Learning and expectations in macroeconomics},
  author={Evans, George W and Honkapohja, Seppo},
  year={2001},
  publisher={Princeton University Press}
}

@article{branch2006restricted,
  title={Restricted perceptions equilibria and learning in macroeconomics},
  author={Branch, William A},
  journal={Post Walrasian Macroeconomics: Beyond the Dynamic Stochastic General Equilibrium Model. Cambridge University Press, New York},
  pages={135--160},
  year={2006}
}

@incollection{hansen2011robustness,
  title={Robustness},
  author={Hansen, Lars Peter and Sargent, Thomas J},
  booktitle={Robustness},
  year={2011},
  publisher={Princeton university press}
}

@article{YAMADA1976294,
title = {Asymptotic behavior of posterior distributions for random processes under incorrect models},
journal = {Journal of Mathematical Analysis and Applications},
volume = {56},
number = {2},
pages = {294-308},
year = {1976},
author = {Keigo Yamada},
}

@book{doob1953stochastic,
  title={Stochastic Processes},
  author={Doob, Joseph L.},
  year={1953},
  publisher={Wiley},
  address={New York}
}

@article{ShaliziEJS2009,
author = {Cosma Rohilla Shalizi},
title = {{Dynamics of Bayesian updating with dependent data and misspecified models}},
volume = {3},
journal = {Electronic Journal of Statistics},
pages = {1039 -- 1074},
year = {2009},
}

@article{FusterEtAlJEP2010,
Author = {Fuster, Andreas and Laibson, David and Mendel, Brock},
Title = {Natural Expectations and Macroeconomic Fluctuations},
Journal = {Journal of Economic Perspectives},
Volume = {24},
Number = {4},
Year = {2010},
Month = {December},
Pages = {67–84}
}

@article{white1982maximum,
  title   = {Maximum likelihood estimation of misspecified models},
  author  = {White, Halbert},
  journal = {Econometrica},
  volume  = {50},
  number  = {1},
  pages   = {1--25},
  year    = {1982},
  publisher = {The Econometric Society}
}

@incollection{newey1994large,
  title        = {Large sample estimation and hypothesis testing},
  author       = {Newey, Whitney K. and McFadden, Daniel},
  booktitle    = {Handbook of Econometrics},
  editor       = {Engle, Robert F. and McFadden, Daniel L.},
  volume       = {4},
  pages        = {2111--2245},
  year         = {1994},
  publisher    = {Elsevier},
  address      = {Amsterdam}
}

@article{Berk1966,
  author  = {Berk, Robert H.},
  title   = {Limiting Behavior of Posterior Distributions When the Model is Incorrect},
  journal = {Annals of Mathematical Statistics},
  year    = {1966},
  volume  = {37},
  number  = {1},
  pages   = {51--58}
}

@article{BunkeMilhaud1998,
  author  = {Bunke, Olaf and Milhaud, Xavier},
  title   = {Asymptotic Behavior of Bayes Estimates Under Possibly Incorrect Models},
  journal = {Annals of Statistics},
  year    = {1998},
  volume  = {26},
  number  = {2},
  pages   = {617--644}
}

@article{KleijnvanderVaart2006,
  author  = {Kleijn, Bas J. K. and van der Vaart, Aad W.},
  title   = {Misspecification in Infinite-Dimensional Bayesian Statistics},
  journal = {Annals of Statistics},
  year    = {2006},
  volume  = {34},
  number  = {2},
  pages   = {837--877}
}

@article{DiaconisFreedman1986,
  author  = {Diaconis, Persi and Freedman, David},
  title   = {On the Consistency of {B}ayes Estimates},
  journal = {Annals of Statistics},
  volume  = {14},
  number  = {1},
  pages   = {1--26},
  year    = {1986}
}

@article{ShenWasserman2001,
  author  = {Shen, Xiaotong and Wasserman, Larry},
  title   = {Rates of Convergence of Posterior Distributions},
  journal = {Annals of Statistics},
  volume  = {29},
  number  = {3},
  pages   = {687--714},
  year    = {2001}
}

@book{GhosalVanDerVaart2017,
  author    = {Ghosal, Subhashis and van der Vaart, Aad W.},
  title     = {Fundamentals of Nonparametric Bayesian Inference},
  publisher = {Cambridge University Press},
  year      = {2017},
  series    = {Cambridge Series in Statistical and Probabilistic Mathematics}
}

@article{fudenberg1993learning,
  title={Learning mixed equilibria},
  author={Fudenberg, Drew and Kreps, David M},
  journal={Games and economic behavior},
  volume={5},
  number={3},
  pages={320--367},
  year={1993},
  publisher={Elsevier}
}

@article{BarronSchervishWasserman1999,
  author  = {Barron, Andrew and Schervish, Mark J. and Wasserman, Larry},
  title   = {The Consistency of Posterior Distributions in Nonparametric Problems},
  journal = {Annals of Statistics},
  volume  = {27},
  number  = {2},
  pages   = {536--561},
  year    = {1999}
}

@article{frick2020misinterpreting,
  title={Misinterpreting others and the fragility of social learning},
  author={Frick, Mira and Iijima, Ryota and Ishii, Yuhta},
  journal={Econometrica},
  volume={88},
  number={6},
  pages={2281--2328},
  year={2020},
  publisher={Wiley Online Library}
}

@article{GhosalGhoshvanderVaart2000,
  author  = {Ghosal, Subhashis and Ghosh, Jayanta K. and van der Vaart, Aad W.},
  title   = {Convergence Rates of Posterior Distributions},
  journal = {Annals of Statistics},
  year    = {2000},
  volume  = {28},
  number  = {2},
  pages   = {500--531}
}

@article{bohren2016informational,
  title={Informational herding with model misspecification},
  author={Bohren, J Aislinn},
  journal={Journal of Economic Theory},
  volume={163},
  pages={222--247},
  year={2016},
  publisher={Elsevier}
}

@article{spiegler2016bayesian,
  title={Bayesian networks and boundedly rational expectations},
  author={Spiegler, Ran},
  journal={The Quarterly Journal of Economics},
  volume={131},
  number={3},
  pages={1243--1290},
  year={2016},
  publisher={Oxford University Press}
}

@article{eyster2005cursed,
  title={Cursed equilibrium},
  author={Eyster, Erik and Rabin, Matthew},
  journal={Econometrica},
  volume={73},
  number={5},
  pages={1623--1672},
  year={2005},
  publisher={Wiley Online Library}
}

@article{molavi2019macroeconomics,
  title={Macroeconomics with learning and misspecification: A general theory and applications},
  author={Molavi, Pooya and others},
  journal={Unpublished manuscript},
  year={2019}
}

@article{ba2023multi,
  title={A multi-agent model of misspecified learning with overconfidence},
  author={Ba, Cuimin and Gindin, Alice},
  journal={Games and Economic Behavior},
  volume={142},
  pages={315--338},
  year={2023},
  publisher={Elsevier}
}

@techreport{murooka2025bayesian,
  title={Bayesian Learning when Players Misspecify Others},
  author={Murooka, Takeshi and Yamamoto, Yuichi},
  year={2025},
  institution={Institute of Social and Economic Research, The University of Osaka}
}

@techreport{murooka2023convergence,
  title={Convergence and Steady-State Analysis under Higher-Order Misspecification},
  author={Murooka, Takeshi and Yamamoto, Yuichi},
  year={2023},
  institution={Osaka School of International Public Policy, Osaka University}
}

@article{schwartzstein2014selective,
  title={Selective attention and learning},
  author={Schwartzstein, Joshua},
  journal={Journal of the European Economic Association},
  volume={12},
  number={6},
  pages={1423--1452},
  year={2014},
  publisher={Oxford University Press}
}

@article{esponda2008behavioral,
  title={Behavioral equilibrium in economies with adverse selection},
  author={Esponda, Ignacio},
  journal={American Economic Review},
  volume={98},
  number={4},
  pages={1269--1291},
  year={2008},
  publisher={American Economic Association}
}

@article{bohren2021learning,
  title={Learning with heterogeneous misspecified models: Characterization and robustness},
  author={Bohren, J Aislinn and Hauser, Daniel N},
  journal={Econometrica},
  volume={89},
  number={6},
  pages={3025--3077},
  year={2021},
  publisher={Wiley Online Library}
}

@article{he2022mislearning,
  title={Mislearning from censored data: The gambler's fallacy and other correlational mistakes in optimal-stopping problems},
  author={He, Kevin},
  journal={Theoretical Economics},
  volume={17},
  number={3},
  pages={1269--1312},
  year={2022},
  publisher={Wiley Online Library}
}

@article{nyarko1991learning,
  title={Learning in mis-specified models and the possibility of cycles},
  author={Nyarko, Yaw},
  journal={Journal of Economic Theory},
  volume={55},
  number={2},
  pages={416--427},
  year={1991},
  publisher={Elsevier}
}

@article{fudenberg2017active,
  title={Active learning with a misspecified prior},
  author={Fudenberg, Drew and Romanyuk, Gleb and Strack, Philipp},
  journal={Theoretical Economics},
  volume={12},
  number={3},
  pages={1155--1189},
  year={2017},
  publisher={Wiley Online Library}
}

@article{frick2023belief,
  title={Belief convergence under misspecified learning: A martingale approach},
  author={Frick, Mira and Iijima, Ryota and Ishii, Yuhta},
  journal={The Review of Economic Studies},
  volume={90},
  number={2},
  pages={781--814},
  year={2023},
  publisher={Oxford University Press US}
}

@article{esponda2016berk,
  title={Berk--Nash equilibrium: A framework for modeling agents with misspecified models},
  author={Esponda, Ignacio and Pouzo, Demian},
  journal={Econometrica},
  volume={84},
  number={3},
  pages={1093--1130},
  year={2016},
}

@article{esponda2021asymptotic,
  title={Asymptotic behavior of Bayesian learners with misspecified models},
  author={Esponda, Ignacio and Pouzo, Demian and Yamamoto, Yuichi},
  journal={Journal of Economic Theory},
  volume={195},
  pages={105260},
  year={2021},
}

@article{esponda2021equilibrium,
  title={Equilibrium in misspecified Markov decision processes},
  author={Esponda, Ignacio and Pouzo, Demian},
  journal={Theoretical Economics},
  volume={16},
  number={2},
  pages={717--757},
  year={2021},
  publisher={Wiley Online Library}
}

@article{heidhues2021convergence,
  title={Convergence in models of misspecified learning},
  author={Heidhues, Paul and K{\H{o}}szegi, Botond and Strack, Philipp},
  journal={Theoretical Economics},
  volume={16},
  number={1},
  pages={73--99},
  year={2021},
  publisher={Wiley Online Library}
}

@article{heidhues2018unrealistic,
  title={Unrealistic expectations and misguided learning},
  author={Heidhues, Paul and K{\H{o}}szegi, Botond and Strack, Philipp},
  journal={Econometrica},
  volume={86},
  number={4},
  pages={1159--1214},
  year={2018},
}

@article{fudenberg2021limit,
  title={Limit points of endogenous misspecified learning},
  author={Fudenberg, Drew and Lanzani, Giacomo and Strack, Philipp},
  journal={Econometrica},
  volume={89},
  number={3},
  pages={1065--1098},
  year={2021},
  publisher={Wiley Online Library}
}

@article{anderson2024existence,
  title={On existence of Berk-Nash equilibria in misspecified Markov decision processes with infinite spaces},
  author={Anderson, Robert M and Duanmu, Haosui and Ghosh, Aniruddha and Khan, M Ali},
  journal={Journal of Economic Theory},
  volume={217},
  pages={105813},
  year={2024},
  publisher={Elsevier}
}

@article{fudenberg2023misspecifications,
  title={Which misspecifications persist?},
  author={Fudenberg, Drew and Lanzani, Giacomo},
  journal={Theoretical Economics},
  volume={18},
  number={3},
  pages={1271--1315},
  year={2023},
  publisher={Wiley Online Library}
}

@article{gagnon2021projection,
  title={Projection of private values in auctions},
  author={Gagnon-Bartsch, Tristan and Pagnozzi, Marco and Rosato, Antonio},
  journal={American Economic Review},
  volume={111},
  number={10},
  pages={3256--3298},
  year={2021},
  publisher={American Economic Association 2014 Broadway, Suite 305, Nashville, TN 37203}
}

@article{madarasz2023projective,
  title={Projective Thinking: Model, Evidence, and Applications},
  author={Madar{\'a}sz, Krist{\'o}f and Danz, David and Wang, Stephanie},
  journal={Evidence, and Applications (May 31, 2023)},
  year={2023}
}

@incollection{kirman1975learning,
  title={Learning by firms about demand conditions},
  author={Kirman, Alan P},
  booktitle={Adaptive economic models},
  pages={137--156},
  year={1975},
  publisher={Elsevier}
}

@techreport{ArrowGreen1973-IMSSS,
  author      = {Arrow, Kenneth J. and Green, Jerry R.},
  title       = {Notes on Expectations Equilibria in Bayesian Settings},
  institution = {Institute for Mathematical Studies in the Social Sciences (IMSSS), Stanford University},
  type        = {Economics Series Working Paper},
  number      = {33},
  address     = {Stanford, CA},
  month       = aug,
  year        = {1973},
}

@unpublished{EspondaPouzo2025-BNR,
  author = {Esponda, Ignacio and Pouzo, Demian},
  title  = {Berk--Nash Rationalizability},
  year   = {2025},
  note   = {Working paper}
}

@article{benaim2005stochastic,
  title={Stochastic approximations and differential inclusions},
  author={Bena{\"\i}m, Michel and Hofbauer, Josef and Sorin, Sylvain},
  journal={SIAM Journal on Control and Optimization},
  volume={44},
  number={1},
  pages={328--348},
  year={2005},
  publisher={SIAM}
}

@article{pemantle1990nonconvergence,
  title={Nonconvergence to unstable points in urn models and stochastic approximations},
  author={Pemantle, Robin},
  journal={The Annals of Probability},
  pages={698--712},
  year={1990},
  publisher={JSTOR}
}

@incollection{Benaim1999DSA,
  author    = {Bena{\"{\i}}m, Michel},
  title     = {Dynamics of Stochastic Approximation Algorithms},
  booktitle = {S{\'e}minaire de Probabilit{\'e}s XXXIII},
  editor    = {Az{\'e}ma, Jacques and {\'E}mery, Michel and Ledoux, Michel and Yor, Marc},
  series    = {Lecture Notes in Mathematics},
  volume    = {1709},
  pages     = {1--68},
  year      = {1999},
  publisher = {Springer},
  address   = {Berlin, Heidelberg},
  doi       = {10.1007/BFb0092699}
}

@book{hahn1973notion,
  title={On the notion of equilibrium in economics: An inaugural lecture},
  author={Hahn, Frank Horace},
  year={1973},
  publisher={Cambridge University Press}
}

@article{eyster2010naive,
  title={Naive herding in rich-information settings},
  author={Eyster, Erik and Rabin, Matthew},
  journal={American economic journal: microeconomics},
  volume={2},
  number={4},
  pages={221--243},
  year={2010},
  publisher={American Economic Association}
}

@article{bohren2025misspecified,
  title={Misspecified Models in Learning and Games},
  author={Bohren, J Aislinn and Hauser, Daniel N},
  journal={Annual Review of Economics},
  volume={17},
  year={2025},
  publisher={Annual Reviews}
}

@article{fudenberg1993self,
  title={Self-confirming equilibrium},
  author={Fudenberg, Drew and Levine, David K},
  journal={Econometrica: Journal of the Econometric Society},
  pages={523--545},
  year={1993},
  publisher={JSTOR}
}

@article{fudenberg1995learning,
  title={Learning in extensive-form games I. Self-confirming equilibria},
  author={Fudenberg, Drew and Kreps, David M},
  journal={Games and Economic Behavior},
  volume={8},
  number={1},
  pages={20--55},
  year={1995},
  publisher={Elsevier}
}

@article{fudenberg1993steady,
  title={Steady state learning and Nash equilibrium},
  author={Fudenberg, Drew and Levine, David K},
  journal={Econometrica: Journal of the Econometric Society},
  pages={547--573},
  year={1993},
  publisher={JSTOR}
}

@book{fudenberg1998theory,
  title={The theory of learning in games},
  author={Fudenberg, Drew and Levine, David K},
  volume={2},
  year={1998},
  publisher={MIT press}
}

@article{jehiel2008revisiting,
  title={Revisiting games of incomplete information with analogy-based expectations},
  author={Jehiel, Philippe and Koessler, Fr{\'e}d{\'e}ric},
  journal={Games and Economic Behavior},
  volume={62},
  number={2},
  pages={533--557},
  year={2008},
  publisher={Elsevier}
}

@article{lanzani2025dynamic,
  title={Dynamic concern for misspecification},
  author={Lanzani, Giacomo},
  journal={Econometrica},
  volume={93},
  number={4},
  pages={1333--1370},
  year={2025},
  publisher={Wiley Online Library}
}

@unpublished{Battigalli1987,
  author = {Battigalli, Pierpaolo},
  title  = {Comportamento razionale ed equilibrio nei giochi e nelle situazioni sociali},
  year   = {1987},
  note   = {Thesis (typescript), Universit{\`a} Bocconi, Milano}
}

@unpublished{fudenberg1988learning,
  title        = {A Theory of Learning, Experimentation and Equilibrium in Games},
  author       = {Fudenberg, Drew and Kreps, David M.},
  year         = {1988},
  note         = {Mimeographed monograph (partial draft 0.11), MIT and Stanford University}
}
